\documentclass{emulateapj}
\usepackage{apjfonts}

\def\ed{\end{document}}

\slugcomment{Accepted for publication in the Astronomical Journal}
\shorttitle{Fornax Kinematics}
\shortauthors{Walker et al.}
\begin{document}
\title{Internal Kinematics of the Fornax Dwarf Spheroidal Galaxy\newline}

\author{Matthew G. Walker\altaffilmark{1}, Mario Mateo\altaffilmark{1}, Edward W. Olszewski\altaffilmark{2}, Rebecca A. Bernstein\altaffilmark{1}, Xiao Wang\altaffilmark{3}, \newline Michael Woodroofe\altaffilmark{4}}

\altaffiltext{1}{Department of Astronomy, University of Michigan, Ann Arbor, MI 48109}
\altaffiltext{2}{Steward Observatory, The University of Arizona, Tucson, AZ, 85721}
\altaffiltext{3}{Department of Mathematics and Statistics, University of Maryland, Baltimore County, Baltimore, MD, 21250}
\altaffiltext{4}{Department of Statistics, University of Michigan, Ann Arbor, MI 48109}


\begin{abstract} 
We present new radial velocity results for 176 stars in the Fornax dwarf spheroidal galaxy, of which at least 156 are probable Fornax members.  We combine with previously published data to obtain a radial velocity sample with 206 stars, of which at least 176 are probable Fornax members.  We detect the hint of rotation about an axis near Fornax's morphological minor axis, although the significance of the rotation signal in the galactic rest frame is sensitive to the adopted value of Fornax's proper motion.  Regardless, the observed stellar kinematics are dominated by random motions, and we do not find kinematic evidence of tidal disruption.  The projected velocity dispersion profile of the binned  data set remains flat over the sampled region, which reaches a maximum angular radius of $65^{\prime}$.  Single-component King models in which mass follows light fail to reproduce the observed flatness of the velocity dispersion profile.  Two-component (luminous plus dark matter) models can reproduce the data, provided that the dark component extends sufficiently beyond the luminous component and the central dark matter density is of the same order as the central luminous density.  These requirements suggest a more massive, darker Fornax than standard core-fitting analyses have previously concluded, with $M/L_V$ over the sampled region reaching 10 to 40 times the $M/L_V$ of the luminous component.  We also apply a non-parametric mass estimation technique, introduced in a companion paper.  Although it is designed to operate on  data sets containing velocities for $>$1000 stars, the estimation yields preliminary results suggesting $M/L_V \sim 15$ inside $r < $1.5 kpc.    
\end{abstract}

\section{Introduction}
\label{sec:intro}

The Milky Way's dwarf spheroidal (dSph) satellite galaxies have stellar masses similar to those of globular clusters ($M_{luminous}\sim10^{6-7} M_{\sun}$), yet they are much more spatially extended ($R \sim 0.5-3$ kpc for dSphs; $R \sim 0.01-0.05$ kpc for globular clusters).   These characteristics give dSphs the lowest luminosity densities of any known galaxies.  The discovery that their internal velocity dispersions all exceed 7 km s$^{-1}$ (Aaronson 1983; Mateo 1998 and references therein) has given rise to competing interpretations and speculations concerning their origin and cosmological significance.  

If dSphs are assumed to be approximately virialized systems, their large velocity dispersions indicate the presence of copious amounts of dark matter.  Estimates of mass-to-light ratios ($M/L$) based on the equilibrium assumption have yielded $M/L \sim 5 - 500$ in solar units for various Milky Way dSphs (Mateo 1998 and references therein; Kleyna et al. 2001; Odenkirchen et al. 2001; Kleyna et al. 2005).  The dSphs are then the smallest, nearest systems believed to reside within dark matter halos, and so provide a convenient and fundamental testing ground for cold dark matter models of structure formation.  

Alternatively, large measured velocity dispersions have been cited as possible evidence that the dSphs are presently undergoing tidal disruption as they orbit within the MW potential (Kuhn 1993; Kroupa 1997; Klessen \& Kroupa 1998; Klessen \& Zhao 2002; Fleck \& Kuhn 2003).  According to this interpretation, dSphs may be in a state far from dynamical equilibrium, and masses derived under that assumption may be inflated.  If the observed stellar velocity dispersions can be attributed to streaming tidal debris projected along the line of sight, the need to invoke dark matter for explaining dSph kinematics subsides, perhaps entirely (Fleck \& Kuhn 2003; Kroupa 1997).  

Large samples of radial velocities measured for dSph stars may be capable of distinguishing between various equilibrium and tidal models by examining the velocity trends across the face of the system.  If a dSph is close to dynamical equilibrium, its stellar motions provide an estimate of the underlying mass distribution.  Tidal disruption is expected to be accompanied by a radial velocity gradient, giving rise to apparent rotation with a characteristic orientation.  We present in this paper new radial velocity results for 176 stars along the line of sight to Fornax.  After combining with previously published results, we test for rotation and then measure the radial velocity dispersion profile extending from the Fornax center to the nominal edge of the luminous component.  We consider the results in the contexts of equilibrium and tidal disruption models.  We also estimate the Fornax mass non-parametrically, applying a technique formally introduced in a companion paper by Wang et al. (2005; hereafter, Paper I).

\section{Observations and Data Reduction}
\label{sec:observations}

\subsection{Photometry, Astrometry, and Target Identification}
\label{subsec:phot}

In order to identify spectroscopic target stars, we first obtained photometric data from 31 fields covering a 110' $\times$ 90' region of sky centered on the Fornax dSph ($\alpha_{2000}$=2:39:52, $\delta_{2000}$=-34:28:09).  These observations took place the nights of 1993 November 30, December 1, and December 10, during photometric conditions.  The data consist of 600-700s exposures in both V and I filters using the 2048 x 2048 TEK 3 CCD detector at the Las Campanas Observatory 40-inch telescope (field size=24'x24', scale=0.7 arcsec pixel$^{-1}$).  The images were processed using twilight flatfield exposures and multiple bias frames.  We used the two-dimensional stellar photometry program DoPHOT (Schechter et al. 1993) for the reductions and placed the resulting instrumental magnitudes on the Kron-Cousins scale (Bessell 1976) using 56 Landoldt photometric standard stars observed during the same nights.  From the formal error values returned by DoPHOT, and multiple measurements of stars in overlapping field regions, we estimate our photometric accuracy to be $\pm 3\%$.  

\begin{figure*}
  \plottwo{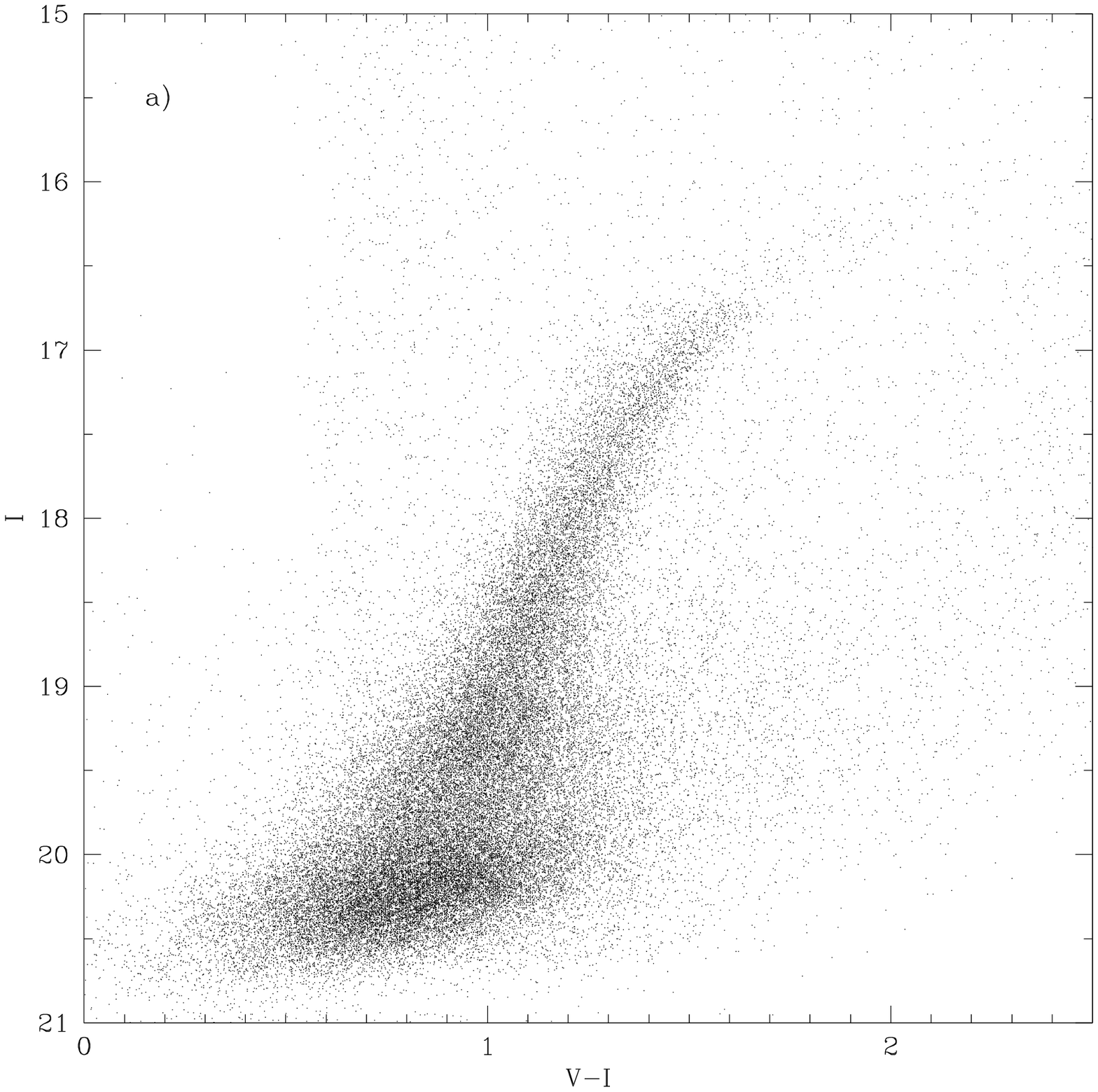}{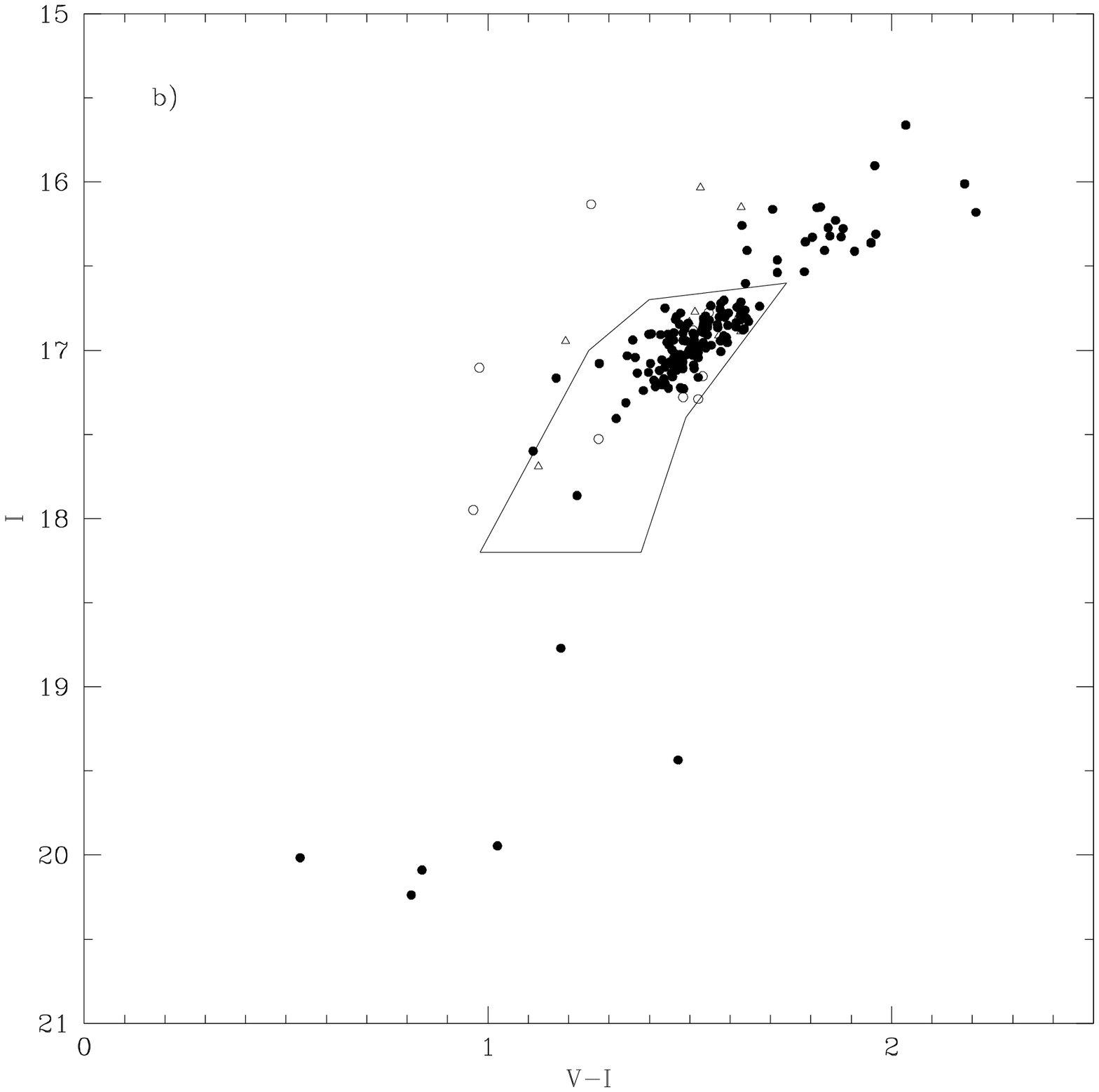}
  \caption{\label{fig:cmdtrue} \scriptsize Fornax red giant branch.  a) includes all stars measured photometrically in a 110' $\times$ 90' region of sky centered on the Fornax dSph.  b) shows only those stars observed spectroscopically, and illustrates the boundaries of our color-magnitude target selection region.  Filled circles are probable Fornax members, based on velocity criteria described in section \ref{subsec:membership}.  Open circles are probable foreground contaminants.  Open triangles represent stars with marginal membership status.  Points located outside the CMD selection region represent stars observed for this study before the photometry comprising this CMD was available, and so were chosen based on the photometry described in Mateo et al. (1991).}
\end{figure*}

The resulting color-magnitude diagram (CMD) is shown in Figure \ref{fig:cmdtrue}, which also shows the region in the CMD from which we selected stars for spectroscopic observation.  The boundaries of this region were chosen to enclose points representing likely Fornax members, as well as those stars bright enough to maintain reasonable integration times during spectroscopy.  The chosen region roughly extends along the brightest $\sim$1.5 I-band magnitudes of the Fornax red giant branch (RGB) and includes more than 4000 Fornax candidate members.  We conservatively chose from among the brighter targets in this selection region for spectroscopic follow-up.  Based on our spectroscopy results, it is feasible to derive accurate radial velocities for stars at even the faint edge of this selection region.

Since one of our spectroscopic observing runs took place prior to 1993, some targets were selected without the benefit of this photometry.  For target identification leading up to the 1992 November-December spectroscopy run, we relied on photometry obtained in 1990 November over a smaller region of sky.  Reduction of these photometric data is described in Mateo et al. (1991; hereafter, M91). Many of these targets fall outside the selection region shown in Figure \ref{fig:cmdtrue}b.  This earlier effort was devoted to observing primarily the brightest candidate members, although several faint RGB stars were selected in order to probe the limits of the instrumentation.  

\begin{figure*}
  \plottwo{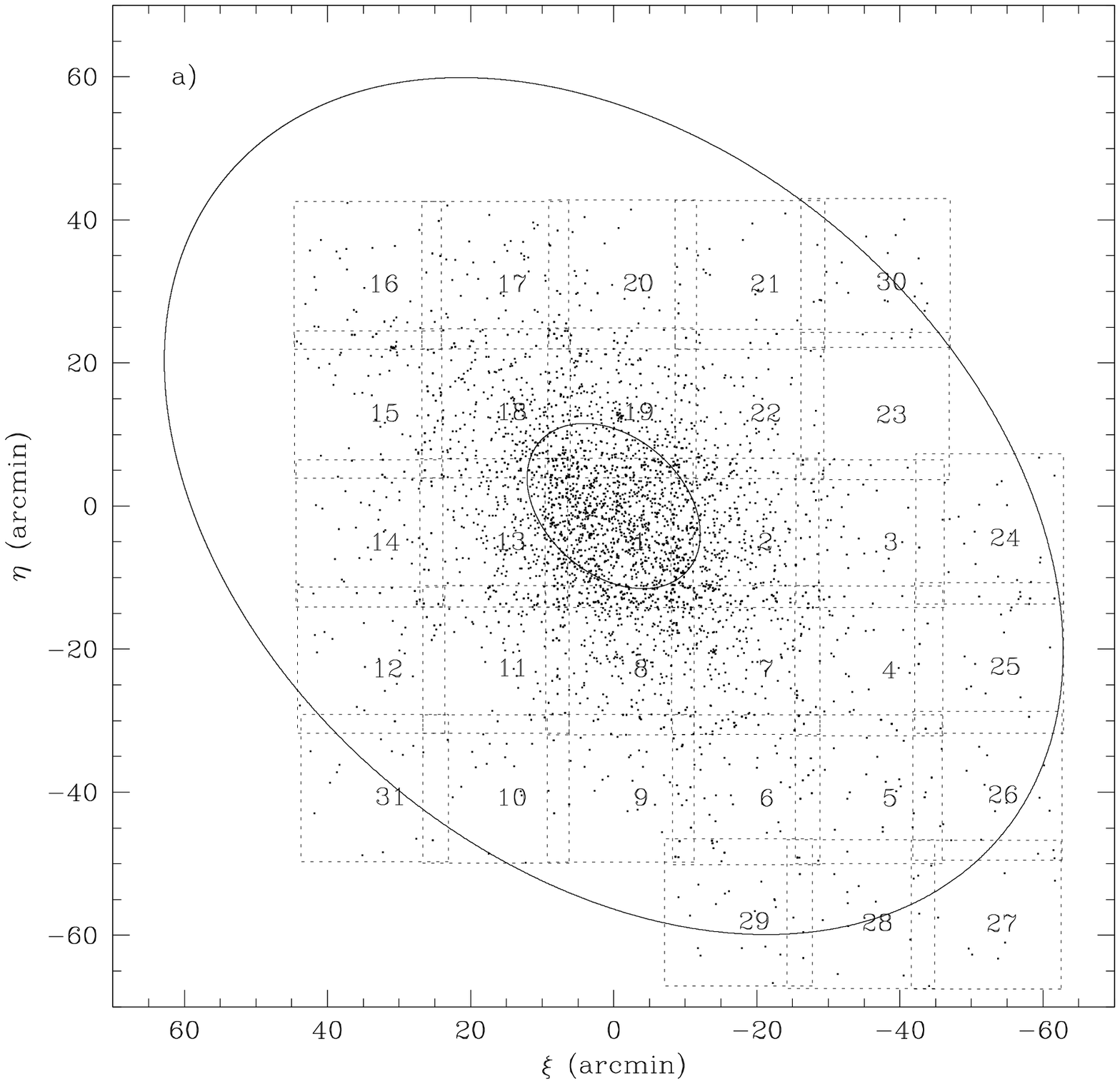}{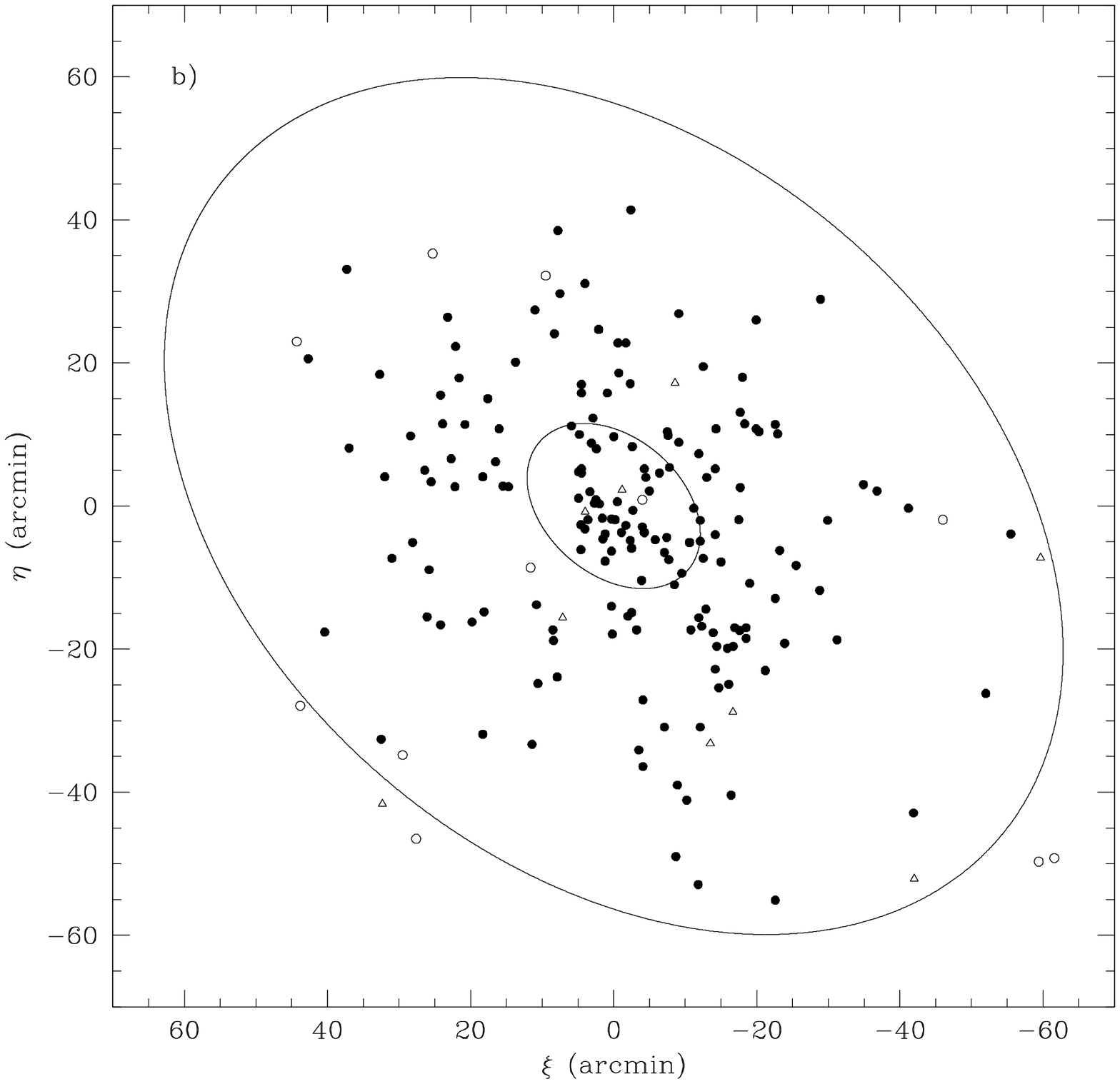}
  \caption{\label{fig:photmap} \scriptsize Maps of (a) all stars meeting the selection criteria discussed in Section \ref{subsec:membership}; overplotted are the boundaries of the 31 photometric fields observed. (b) maps stars for which we measured radial velocities.  Filled circles represent stars later determined to be probable Fornax members.  Open circles represent stars rejected as probable foreground contaminants on the basis of their radial velocities.  Open triangles represent stars having velocities marginally consistent with Fornax membership (see section \ref{subsec:membership}).  The inner and outer ellipses are the King core and tidal radii, respectively, which have published semi-major axis values $r_{core}=13.7\arcmin \pm 1.2\arcmin$ and $r_{tide}=71.1\arcmin \pm 4.0\arcmin$, with ellipticity $\epsilon=0.30\pm0.01$ (Irwin $\&$ Hatzidimitriou 1995).  The standard coordinate system is centered on the Fornax dSph such that ($\xi$,$\eta$)=(0,0) corresponds to $\alpha_{2000}$=2:39:52, $\delta_{2000}$=-34:28:09.  North is toward the top of the figure, east is to the left.}
\end{figure*}

An additional factor entering our target selection was a star's sky position relative to the center of Fornax.  To aid this selection we converted the (x,y) CCD position returned by the DoPHOT centroid algorithm into equatorial coordinates using the IRAF routines TFINDER and CCTRANS, and tied our astrometry to the USNO-1B system using up to several hundred USNO stars per CCD frame.  From measurements of stars in overlapping fields, we estimate the $2\sigma$ astrometric accuracy to be better than $0.2\arcsec$.  In deciding on eventual spectroscopic targets, a selection routine closely following the stellar density distribution is inadequate.  The outer, sparsely populated regions are of disproportionately high kinematic interest.  Nevertheless, we wished to obtain a large sample with a high fraction of Fornax members.  In the end we chose at least 2-3 candidate members in all the outer Fornax CCD fields, and limited our selection in the inner fields to $\sim$ 5-8 stars per field.  Figure \ref{fig:photmap} maps the locations of the stars falling within our CMD selection region, and identifies which of those stars we ultimately observed spectroscopically.  

\subsection{Spectroscopy}
\label{subsec:spectroscopy}

We obtained spectra of the selected red giants over four observing runs occurring 1992 November 28 - December 7, 1993 December 12-20, 1994 October 22-28, and 2002 December 12-15.  The 1992-1994 spectra were taken at the Las Campanas Observatory 2.5m du Pont Telescope equipped with an echelle spectrograph and 2D-Frutti photon counting detector (Shectman 1984).  The 2002 spectra were acquired at the Magellan 6.5m Clay Telescope with the Magellan Inamori Kyocera Echelle (MIKE) spectrograph (Bernstein et al. 2003) using a 1.0'' $\times$ 5.0'' slit and MIKE's red-side CCD detector set to obtain spectra between $4400-8000$ \AA.  As they came from different telescopes/instruments, the 1992-1994 spectra and 2002 spectra were reduced independently.  We followed the same general procedure in both cases.  In what follows, we describe the observation and reduction procedure specific to the 2002 spectra.  Where details differ regarding the 1992-1994 spectra, we comment parenthetically.   

Object exposure times ranged from 360-720 seconds (900-2400s for the 1992-1994 runs), with most at 600s (1800s).  In addition, we took 1s (360s) exposures of a Th-Ar comparison arc lamp before and after object exposures.  Other exposures included quartz-illuminated dome flats and spectra of bright radial velocity standard stars used for aperture identification and velocity calibration.  We used the IRAF data reduction software package to reduce the raw spectra to heliocentric radial velocities.  After overscan and bias subtraction, we produced a master flatfield frame for each night by averaging quartz-illuminated dome exposures.  Flatfield frames were then normalized using the IRAF task APFLATTEN, which models and removes both the spatial profile and spectral shape of the illumination pattern, leaving only the sensitivity variations.  We corrected our object spectra for these sensitivity variations by dividing each by the appropriate normalized flatfield frame.  We then ran the IRAF task APALL (with FORMAT keyword set to ``strip'') on the spectra of the bright standard stars to obtain two-dimensional traces of the echelle orders on the detector and rectify the spectra.  We found the order location on the detector to remain quite stable over the course of the run (MIKE is fixed with respect to gravity).  We then ran APALL on all the target object spectra to obtain a rectified spectrum from each order, referencing the trace pattern identified for the most recently observed bright star.  Thorium-argon comparison spectra were rectified in exactly the same manner as the individual stellar spectra they would eventually correct and calibrate.

For the 2002 data we had also to remove cosmic rays (the 1992-1994 data obtained via photon-counting device did not suffer from cosmic rays).  As a first pass, we ran the IRAF task COSMICRAYS, specifying conservative thresholds so as to remove only the most conspicuous events.  The majority of cosmic rays were then removed by the task CONTINUUM, using a 10th-order cubic spline to replace any pixel value above an upper sigma threshold determined by eye to optimize accurate cosmic ray identification.  

We then employed the tasks IDENTIFY and REIDENTIFY to convert the Th-Ar spectra from pixels to wavelength space.  A typical arc lamp spectrum would have, for the ten orders (four orders) we eventually used for velocity measurement, 140 lines (320 lines) reidentified with an rms scatter of 0.09 \AA (0.03 \AA) for a fourth-order polynomial fit to the wavelength solution.  These wavelength calibrations were then applied to the object spectra using the DISPCOR task, which converts units in the dispersion direction from pixels to Angstroms using the weighted solutions to the two nearest comparison arc spectra as references.  

Up to this point, the spectra remained two-dimensional, and we had treated each row in the spatial direction separately.  This is necessary because the spatial and spectral axes are not orthogonal within each order (i.e., the spectral lines are tilted by $\sim$ 20$^{\circ}$ with respect to the spatial direction on the detector).  By wavelength-calibrating each row in the spatial direction explicitly, we eliminate this problem and retain the full spectral resolution of the instrument.  

We then converted the wavelength-calibrated two-dimensional spectra to one-dimensional spectra (for the 1992-1994 spectra, this was accomplished by the ECDISPCOR subroutine).  For the 2002 spectra, first we used SCOMBINE to sum the five spatial rows at the center of each spectral order, which we had determined to carry the stellar signal.  Separately, we averaged two rows located sufficiently far from that aperture center so as to identify primarily the sky spectrum.  Finally, we used SCOMBINE to subtract the normalized sky spectrum from the summed stellar spectrum of the same aperture.  A second pass with CONTINUUM then fitted and subtracted the continuum signal, and IMREPLACE then replaced with zero any pixel with absolute value greater than fifty.  Our velocity measurements are limited to the wavelength range $\sim 4900-5600$ \AA ($\sim 5120-5460$ \AA), to improve precision of the results.  The wavelength range chosen for the 1992-1994 data corresponds to the four echelle orders determined to yield the most precise velocity measurements from these spectra.  At wavelengths redder than this range, the metal-poor spectra provide few absorption lines, while at wavelengths blueward of $4900$ \AA, there is little continuum flux from the red giant targets.  The strongest absorption lines within the selected region belong to the magnesium-B triplet having a rest wavelength near $5170$ \AA, while the many weaker absorption lines contribute usefully in aggregate to the cross correlation function.

We calculated heliocentric radial velocities using the cross-correlation package FXCOR.  We cross-correlated the extracted spectrum of each target star against a high signal-to-noise ratio template consisting of the sum of 27 (75) spectra of bright radial-velocity standard stars.  The spectrum for each standard had first been shifted to a common heliocentric redshift equal to that of the star HD6655 (v$_{helio}$=$19.5 \pm 0.3$ km s$^{-1}$; Udry et al. 1999).  Prior to cross-correlation, we filtered the 10970 (8192)-pixel spectra with a ramp function, cutting on at 175 (100) wavenumbers, increasing linearly to full value at 200 (170) and then decreasing linearly from 2200 (700) to a cutoff at 2500 (1000).  A Gaussian fit then located the center of the cross-correlation peak, thereby specifying the radial velocity difference between the object and template.
  
\subsection{Measurement Uncertainties}
\label{subsec:uncertainties}

As a check on the reliability of the extractions and cross-correlations, we compare independent velocity results obtained for the bright standard stars observed multiple times.  Let $N_{bright}$ be the number of standard stars observed, and let $N_b$ be the number of independent observations of standard star $b$.  Letting $N_B$ be the total number of individual measurements accumulated for standard stars ($N_B=\sum_{b=1}^{N_{bright}}N_b$), we then define the cumulative sample variance over all independent measurements of such stars to be $\sigma_{bright}^2=(N_B-N_{bright})^{-1}\sum_{b=1}^{N_{bright}}\sum_{j=1}^{N_b}(v_{b,j}-\langle v \rangle_b)^2$.  For our sample we find $\sigma_{bright}=0.89$ km s$^{-1}$ ($\sigma_{bright}=0.72$ km s$^{-1}$) for $N_B$=24 ($N_B$=107).  This indicates a satisfactory level of internal consistency for our purposes.

To calculate the internal measurement uncertainty, $\sigma_j$, associated with each independent velocity measurement, $v_j$, we assume that multiple measurements of a given star having true velocity $v_{true}$ follow a Gaussian distribution with mean $v_{true}$ and variance $\sigma_j^2$.  Multiple measurements will be distributed as $v_{true}+\sigma_j\epsilon_j$, where $\epsilon_j$ is a random variable fixed by measurement and following a Gaussian probability distribution with mean zero and variance unity (a standard normal distribution).  For simplicity we estimate $v_{true,i}$, the true velocity of star $i$, from $N_i$ independent measurements as\footnote{We follow the convention by which the estimation of quantity $q$ is denoted $\hat{q}$.} $\hat{v}_{true,i}=N_i^{-1}\sum_{j=1}^{N_i}v_{ij}$.  We make the further assumption that the difference $\hat{v}_{true,i}-v_{true,i}$ is negligible.  This assumption is perhaps naive; however, we find for this  data set that a rigorous treatment, properly considering the uncertainty in $\hat{v}_{true}$, gives nearly identical results.  With these assumptions we then express the $j^{th}$ velocity measurement of star $i$ as 
\begin{equation}
  \label{eq:errors}
  v_{ij}=\hat{v}_{true,i}+\sigma_{ij}\epsilon_{ij},
\end{equation}

We model the $\sigma_{ij}$ as a function of the Tonry-Davis $R$ value (Tonry \& Davis 1979), which FXCOR calculates as the ratio of the selected cross-correlation peak height to the average height of the nonselected peaks.  We express the relationship as 
\begin{equation}
\label{eq:tdr}
\sigma_{ij}=\frac{\alpha}{(1+R_{ij})^{x}}.
\end{equation}
This two-parameter model generalizes the original Tonry \& Davis formalism, which assumed $x=1$.  We find that we are better able to reproduce the empirical sample variances obtained from repeat measurements by treating $x$ as a free parameter. 

The base-10 logarithm of the squared error in the $i^{th}$ measurement is then
\begin{equation}
  \label{eq:errorsub}
  \log[(v_{ij}-\hat{v}_{true,i})^2]=2\log\alpha-2x\log(1+R_{ij})+\log(\epsilon_{ij}^2).
\end{equation}
The term $\log(\epsilon_{ij}^2)$ has mean $\langle\log(\epsilon_{ij}^2)\rangle=-0.55$, from Monte Carlo simulations.  If we define $\delta_{ij}\equiv\log(\epsilon_{ij}^2)+0.55$, Equation \ref{eq:errorsub} becomes
\begin{equation}
  \label{eq:delta}
  \log[(v_{ij}-\hat{v}_{true,i})^2]=2\log\alpha-2x\log(1+R_{ij})+\delta_{ij}-0.55.
\end{equation}
  
We then estimate ${x}$ and $\alpha$ by linear regression using the ($R_{ij},v_{ij}$) data from only those stars with repeat measurements, and recognizing that the $\delta_{ij}$ have mean value $\langle\delta_{ij}\rangle=0$.  Including bright standard stars, our 2002 data contain 25 (139 for the 1992-1994 data) repeat observations of 6 (19) different stars.  From the 1992-1994 data we obtain the estimates $\hat{\alpha}=6.0$ km s$^{-1}$ and $\hat{x}=0.50$.  Because the 2002 observations contain fewer repeats, and no repeat observations of low-$R$ target stars, we adopt $x=0.50$ for the 2002 data and then estimate $\hat{\alpha}$ with a least-squares fit to find $\alpha=7.6$ km s$^{-1}$ for the 2002 results.  Using the appropriate values for the parameters $\alpha$ and $x$, we then calculate the uncertainty in each individual velocity measurement using Equation \ref{eq:tdr}. 

Within the 1992-1994 data, we were able to check the stability of our velocity zero-point both night-to-night and run-to-run using individual spectra obtained each night for the bright star CPD-35$^{\circ}$919, located just west of the Fornax center.  Within a given run, the night-to-night scatter in the measured velocity of CPD-35$^{\circ}$919 is nearly identical to its estimated internal errors.  Comparing the mean velocity measured for this star in each run, we find no significant run-to-run scatter.   

Since the 1992-1994 and 2002 spectra were cross-correlated using different template spectra, we searched for any systematic zero-point velocity difference between the two independent  data sets.  To accomplish this, we cross-correlated the target spectra from the 2002 run against the radial velocity template used for the 1992-1994 data.  We found the resulting measured velocities to differ by less than 0.05 km s$^{-1}$ with respect to their values derived from the 1992-1994 velocity template.  Finding an equally small discrepancy when cross-correlating the 1992-1994 target spectra against the 2002 template, we take any zero-point velocity offset between the two  data sets to be insignificant.  To measure the overall zero-point offset of the entire combined  data set, we considered the measured velocities of all bright radial velocity standard stars separately, finding a mean discrepancy of $\langle v_{observed}-v_{published}\rangle$ =1.25 km s$^{-1}$, which we take to be our zero-point error.  Subtracting this value from all our velocity results, we then measure a mean value of $-0.3 \pm 1.2$ km s$^{-1}$ for eight twilight sky spectra (corrected individually for diurnal and annual motions).  Thus we are confident that we have placed the velocities on a true heliocentric zero point.

\section{Results}
\label{sec:results}
\subsection{Heliocentric Radial Velocities}
\label{sec:velocities}

In Table \ref{tab:results} (tables are located after the appendix) we present the heliocentric radial velocities and formal uncertainties derived from each individual observation.  Entries are sorted by date of first observation, with any repeat measurements of the same star listed directly below.  Additional information includes the equatorial coordinates, date and time of observation, and the measured $I$,$V-I$ photometry for each object.  The distance $R$ is the angular distance between the center of Fornax, taken from Irwin \& Hatzidimitriou (1995; hereafter, IH95) to be $\alpha_{2000}$=02:39:52.3, $\delta_{2000}$=--34:28:09.0, and the projection of the radial position vector on the plane centered on these coordinates.  The position angles are defined to be 0$^{\circ}$ due north at the tangent point and 90$^{\circ}$ due east.  The final column indicates whether we judge the star to be a probable member (``Y'') or nonmember (``N'') of Fornax, based on photometric and velocity criteria (section \ref{subsec:membership}).  Several stars present borderline cases for membership, and we mark their membership status as ``?''.  If the membership status is other than ``Y,'' the superscript indicates whether this is due to the star's photometry (``p''), velocity (``v''), or both (``v,p'').  We include all Fornax targets, as well as the bright star CPD-35$^{\circ}$919.  Table \ref{tab:standards} lists the radial velocity results for the observed standards.    


We observed several stars on multiple occasions.  In subsequent analyses we take the heliocentric radial velocity of each multiply-observed star to be the average of that star's individual velocity measurements weighted by their respective uncertainites.  Table \ref{tab:repeats} gives the weighted mean velocity, $\chi^2$ value, and the probability, $p(\chi^2)$, for stars having multiple measurements.  Several of the Fornax stars in our sample have velocities previously published by M91.  Table \ref{tab:compare} compares our measurements for these stars with those of M91.  For 11 of the 14 stars common to both  data sets, we find agreement to within the measurement uncertainties.  Of the remaining three, two have velocities reported by M91 differing from our measurements by $\sim2.5\sigma$, while the third differs by $\sim10\sigma$.  See section \ref{sec:discussion} for a discussion on velocity variability and its effects on our results.


\subsection{Fornax Membership}
\label{subsec:membership}

\begin{figure*}
  \plotone{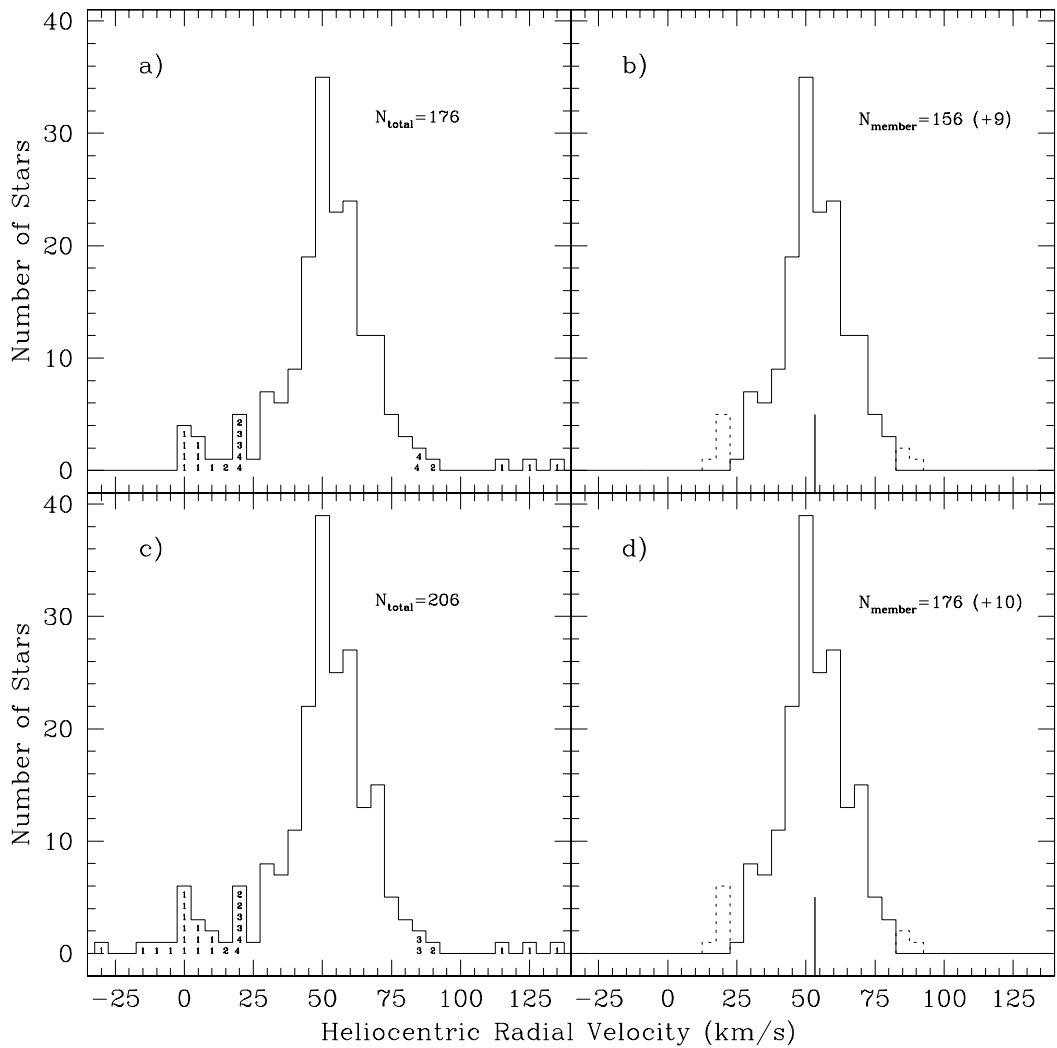}
  \caption{\label{fig:hist} \scriptsize Heliocentric radial Velocity distribution of a) all 176 Fornax candidate member stars whose velocities are presented in Table \ref{tab:results}.  Stars that were later rejected by an iterative membership determination algorithm are numbered according to which iteration rejected them (e.g., ``1''= 1st iteration); b) the 156 stars determined to be probable Fornax members; c) all 209 Fornax candidate member stars after combining our data with that of M91.  Again, numbers specify which iteration removed probable nonmembers; d) the 176 probable Fornax members from the combined  data set.  In (b) and (d), a thick vertical line marks the mean velocity of members calculated using maximum likelihood statistics.  The regions enclosed by dotted lines in (b) and (d) represent those stars rejected in iterations 2,3 and 4.  We consider these to be borderline members.}
\end{figure*}

We identify and exclude from our sample those stars that are likely to be foreground contaminants.  Having passed positional and photometric criteria, these interlopers are best identified as outliers in the observed velocity distribution.  The heliocentric radial velocity of Fornax is $\sim$53 km s$^{-1}$, so the velocity distribution of its stars overlaps that of foreground stars near $v \sim 0$ km s$^{-1}$.  This is apparent in Figure \ref{fig:hist}a, which depicts the distribution of the radial velocities listed in Table \ref{tab:results}.  As the derived mass of a pressure-supported system scales as the square of velocity dispersion, it is imperative that we obtain a sample with minimal contamination from non-members.  In order to accomplish this objectively, we adopt the robust bi-weight estimator (Beers et al. 1990) which determines a characteristic distribution width, $\sigma_{bw}$, equal to the standard deviation in the special case of a normal distribution.  Since 99\% of the members in a normally distributed sample are located within 2.58$\sigma$ of the mean, we select as a membership criterion $|v_i-\langle v \rangle|<2.58\sigma_{bw}$, and iteratively remove those stars failing to satisfy this condition.  This rejection process converges after four iterations, identifying 20 stars as probable foreground.  We are left with a sample of 156 new stars we consider to be members of the Fornax dSph.  Their radial velocity distribution is shown in Figure \ref{fig:hist}b.

On examination of Figure \ref{fig:hist}a, which specifies the iteration that removes each of the rejected stars, one may reasonably wonder whether iterations 4, 3, and possibly 2 of the rejection algorithm remove what are actually Fornax member stars.  The eye is tempted to include the three stars with radial velocities in the range 82.5-92.5 km s$^{-1}$ in the wing of the distribution centered on Fornax's systemic velocity.  Reinstating these stars as probable Fornax members and then forcing symmetry on the overall Fornax distribution would argue for the additional reinstatement of the rejected stars falling in the 12.5-22.5 km s$^{-1}$ range.  Where it is practical, we examine the effects on our results of retaining stars rejected in the second, third and fourth iterations, and note that the true membership of Fornax probably includes some, but perhaps not all, of these stars.  Pending a larger  data set, we leave the membership status of these borderline cases an open question.  

Hereafter we combine our new velocity results with the previously published sample of M91.  The 44 stellar velocities measured by M91 were drawn from stars belonging to one of two distinct fields: one centered on the Fornax core, another located along the major axis 25 arcmin southwest of the center.  We recalculate weighted mean velocities for any stars measured multiple times and/or in both  data sets, using the quoted uncertainties for each observation.  In Figure \ref{fig:hist}c we show the radial velocity distribution of this combined  data set, now consisting of velocities for 206 stars.  The algorithm described above for membership determination then rejects (again converging after four iterations) 28 stars as probable foreground, including all 20 of the stars that had been rejected before the addition of the M91 data.  The eight additional rejected stars come from M91 alone, and were rejected in that study as well.  Figure \ref{fig:hist}d shows the velocity distribution of the 176 stars retained as probable Fornax members.  Again, we examine the effect of retaining those stars rejected in iterations three and four (giving a sample with N=182 members), as well as iterations two, three, and four (giving N=186 members).

\subsection{Rotation and the Proper Motion of Fornax}
\label{subsec:rotation}

The relative motion between the Sun and Fornax contributes a velocity component, $v_{rel}(l,b)$, along the line of sight to each Fornax star.  A given star's galactic rest frame (GRF) radial velocity, $v_{r,GRF}$, is related to its heliocentric rest frame (HRF) velocity\footnote{In either rest frame, a star's ``radial'' velocity is the velocity along the line of sight from the Sun to the star.  GRF radial velocities are defined as those measured by an observer at the Sun's location, but at rest with respect to Fornax.}, $v_{r,HRF}$, by
\begin{equation}
  \label{eq:grf}
  v_{r,GRF}=v_{r,HRF}+v_{rel}(l,b).
\end{equation}
A gradient in $v_{rel}(l,b)$ across the face of an object as large in solid angle as Fornax will tend to produce a spurious gradient in the HRF radial velocities (see Lin \& Dong 2006, \textit{in prep}).  A non-rotating object might thereby give the appearance of rotation to the HRF observer, and a truly rotating object may appear to rotate at a different speed and/or about a different axis.  In order to test for Fornax rotation we correct for this perspective effect by placing our HRF radial velocity  data set in the GRF.

Let $\mathbf{v}_*$ be the GRF space velocity of a given Fornax star; we seek to determine the component of $\mathbf{v}_*$ along the line of sight from the Sun to the star.  Let $\mathbf{v}_{\sun}$ be the velocity of the Sun with respect to the LSR, and let $\mathbf{v}_{F}$ be the bulk velocity of Fornax with respect to the LSR.  Then the projection of the relative motion between the Sun and Fornax along the line of sight to the given star is the sum of scalar products:
\begin{equation}
  \label{eq:vrel}
  v_{rel}(l,b)=\frac{\bold{v}_*}{|\bold{v}_*|}\cdot(\bold{v}_{\sun}-\bold{v}_{F}).
\end{equation}
We apply Equations \ref{eq:grf} and \ref{eq:vrel} along the line of sight to every star in the Fornax velocity  data set in order to determine each star's GRF radial velocity.  For each star's GRF radial velocity uncertainty we adopt the corresponding measurement uncertainty originally estimated in Section \ref{subsec:uncertainties}.  We adopt the values $\bold{v}_{\sun}=13.7$ km s$^{-1}$ toward ($l,b$)=(26.6$^{\circ}$,31.4$^{\circ}$) (Dehnen \& Binney 1998).  The three components of Fornax's velocity with respect to the LSR are computed from Fornax's heliocentric radial velocity and proper motion, via Equations 44 - 46 of Piatek et al. (2002).  We adopt +53.3 km s$^{-1}$ as the heliocentric radial velocity of Fornax.  Piatek et al. (2002) and Dinescu et al. (2004) provide independent measurements of the Fornax proper motion.  Since their results agree only at the $\sim 2\sigma$ level we consider both cases independently, giving two possible GRF radial velocity  data sets.  

\begin{figure*}
  \plotone{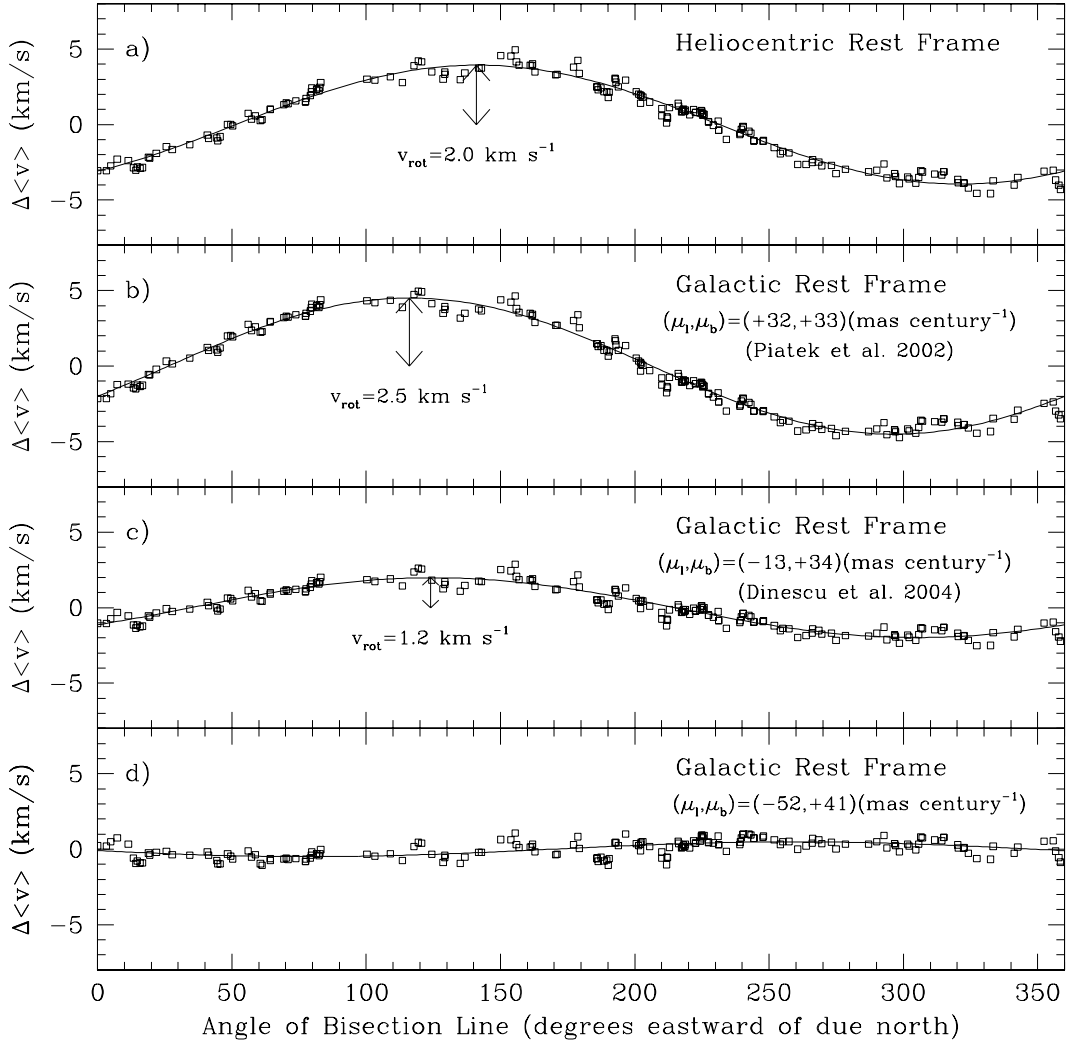}
  \caption{\label{fig:sinfit} \scriptsize Rotation signal of Fornax.  The difference between the radial velocity of Fornax members on either side of a line passing through Fornax's center is plotted as a function of the position angle of that line.  a) computed using the measured heliocentric rest frame radial velocities, uncorrected for perspective-induced rotation (see Section \ref{subsec:rotation}).  b) computed from GRF radial velocities obtained using the Fornax proper motion measurement of Piatek et al. (2002).  c) computed from GRF radial velocities obtained using the Fornax proper motion measurement of Dinescu et al. (2004).  d) computed from GRF radial velocities obtained under the assumption that Fornax does not rotate.}
\end{figure*}

We then test the two resulting GRF radial velocity data sets for rotation.  In both cases we consider the position angle of every star in the intermediate N=182 member sample to coincide with that of a prospective rotation axis.  For each star we bisect the face of Fornax with a line having that star's position angle, and then calculate the mean GRF radial velocity from the member stars on either side of the line.  Figure \ref{fig:sinfit} plots the hemispheric mean velocity \textit{difference} as a function of the bisecting line's position angle.  Panels (b) and (c) of Figure \ref{fig:sinfit} depict the GRF rotation signal assuming the Piatek et al. and Dinescu et al. proper motions, respectively.  For comparison, panel (a) shows the HRF apparent rotation signal, uncorrected for any perspective-induced velocity gradient.  The half-amplitude of the sinusoid fit in each plot measures a charactersistic rotation speed, whereas the sinusoid's phase indicates the orientation of the rotation axis.  We summarize the results of this test in Table \ref{tab:rotation}.  Columns 4 and 5 of Table \ref{tab:rotation} list the characteristic rotation speed and orientation of the rotation axis, respectively.  The uncertainties given for these values reflect the range of values obtained using all proper motions allowed within the published ($1\sigma$) proper motion uncertainties.  Uncertainties in the solar motion and in the HRF radial velocities of the Fornax stars are not considered here.  To assess the significance of a rotation detection, we performed Monte Carlo simulations in which $10^4$ samples of 182 stars having positions of those in the actual sample were drawn at random from a non-rotating, Gaussian velocity distribution with $\sigma=12.4$ km s$^{-1}$.  Column 6 of Table \ref{tab:rotation} gives the percentage of these artificial samples for which we would measure a rotation speed greater than the observed speed listed in column 4.  A lower percentage indicates a more statistically significant observed rotation.

We find that although both published proper motion measurements imply Fornax rotation about an axis at position angle $\sim 115^{\circ}$, only the rotation detected using the Piatek et al. (2002) proper motion is (marginally) statistically significant.  Fewer than $10\%$ of Monte Carlo trials produce rotation as fast as the $\sim2.5$ km s$^{-1}$ implied by the Piatek et al. (2002) proper motion.  Nearly three in four trials produce the $\sim1.2$ km s$^{-1}$ rotation implied by the Dinescu et al. (2004) proper motion.  Therefore, due to perspective effects and the existing uncertainty in the proper motion, we cannot state definitively how or even if Fornax rotates.  If we simply \textit{assume} that Fornax does not rotate, we can use the apparent rotation seen in the HRF radial velocity data indirectly to ``measure'' Fornax's proper motion.  The uncorrected HRF data indicate $\sim 2.0$ km s$^{-1}$ ``rotation'' about an axis at $\sim 140^{\circ}$.  A Fornax GRF proper motion of ($\mu_l,\mu_b$)=($-52,+41$)(mas century$^{-1}$), when applied to these data, would produce a GRF radial velocity  data set showing zero rotation (panel (d) of Figure \ref{fig:sinfit}).  

These results and Fornax's velocity dispersion of $> 10$ km s$^{-1}$ (Section \ref{subsec:dispersion}) indicate that, aside from a possible tidal interpretation (Section \ref{subsec:tides}), any real rotational component is dynamically insignificant.  Given the proper motion ambiguity, we use the HRF radial velocity values of Table \ref{tab:results} in the velocity dispersion calculations that follow.  We demonstrate in Section \ref{subsec:dispersion} that velocity dispersions measured in the HRF differ negligibly from their plausible GRF values.

\subsection{Velocity Dispersion}
\label{subsec:dispersion}

\subsubsection{Maximum Likelihood}
We use maximum likelihood statistics to estimate the mean heliocentric velocity and intrinsic velocity dispersion of those stars we have determined to be members.  Let $v_i$, $u_i$, and $\sigma_i$ now be the measured radial velocity, the true radial velocity, and the internal measurement uncertainty, respectively, for the $i^{th}$ of $N$ stars.  Then $v_i=u_i+\sigma_i\epsilon_i$, where the values $\{\epsilon_1$, ..., $\epsilon_N\}$ have a standard normal probability distribution.  There are two sources of variability in $v_i$: the random, internal measurement uncertainty, $\sigma_i$, and the intrinsic radial velocity dispersion, denoted $\sigma_p$, for the stars in the sample.  The latter is the physical quantity of interest.  If we assume that the values $\{v_1, ..., v_N\}$ have a Gaussian distribution centered on the mean true velocity, denoted $\langle u \rangle$, then their joint probability function is the product of their individual Gaussian probabilities: 
\begin{equation}
  \label{eq:jointGaussian}
p(\{v_1, ..., v_N\})=\displaystyle\prod_{i=1}^{N}\frac{1}{\sqrt{2\pi(\sigma_i^2+\sigma_p^2)}}\exp\biggl[-\frac{1}{2}\frac{(v_i-\langle u \rangle)^2}{(\sigma_i^2+\sigma_p^2)}\biggr].
\end{equation}
Estimates of $\langle u \rangle$ and $\sigma_p$, denoted $\hat{\langle u \rangle}$ and $\hat{\sigma_p}$, are determined numerically as the values that maximize the natural logarithm of the probability function, 
\begin{equation}
  \label{eq:log}
\ln(p)=-\frac{1}{2}\sum_{i=1}^{N}\ln(\sigma_i^2+\sigma_p^2)-\frac{1}{2}\sum_{i=1}^N\frac{(v_i-\langle u \rangle)^2}{(\sigma_i^2+\sigma_p^2)}-\frac{N}{2}\ln(2\pi).
\end{equation}
As the logarithm is a monotonic function, this is equivalent to maximizing $p$ itself (Rice 1995).  To estimate confidence intervals for $\hat{\langle u \rangle}$ and $\hat{\sigma_p}$ we recognize that the Gaussian probability distributions for $(\hat{\langle u \rangle}-\langle u \rangle)$ and $(\hat{\sigma_p}-\sigma_p)$ have centers at zero and a joint variability described by a covariance matrix.  This covariance matrix, $A$, has elements
\begin{equation}
  \label{eq:covariance}
A=
   \left(\begin{array}{cc}
    a & c\\
    c & b
   \end{array}\right),
\end{equation}
where diagonal elements $a$ and $b$ are the variances of $\langle u \rangle$ and $\sigma_p$, respectively.  We determine $a$ and $b$ from the inverse of the covariance matrix, which has the property
\begin{equation}
  \label{eq:inverse}
  A^{-1}=
   \left(\begin{array}{cc}
    \frac{\partial^2\ln(p)}{\partial\langle u \rangle^2}\biggr|_{\tiny{(\hat{\langle u \rangle},\hat{\sigma_p})}} & \frac{\partial^2\ln(p)}{\partial\sigma_p\partial\langle u \rangle}\biggr|_{\tiny{(\hat{\langle u \rangle},\hat{\sigma_p})}}\\
    \frac{\partial^2\ln(p)}{\partial\langle u \rangle\partial\sigma_p}\biggr|_{\tiny{(\hat{\langle u \rangle},\hat{\sigma_p})}} & \frac{\partial^2\ln(p)}{\partial\sigma_p^2}\biggr|_{\tiny{(\hat{\langle u \rangle},\hat{\sigma_p})}}
   \end{array}\right).
\end{equation}
Let $Z_{\alpha/2}$ denote the $\frac{\alpha}{2}$ quantile of the standard normal distribution.  For confidence intervals containing the physical values $\langle u \rangle$ and $\sigma_p$ with $100(1-\alpha)\%$ probability, we report the mean velocity and velocity dispersion as $\hat{\langle u \rangle} \pm Z_{\alpha/2}\sqrt{a}$ and $\hat{\sigma_p} \pm Z_{\alpha/2}\sqrt{b}$, respectively.  Conventional 68\% confidence intervals are given by $\hat{\langle u \rangle} \pm \sqrt{a}$ and $\hat{\sigma_p} \pm \sqrt{b}$.

We estimate the mean true velocity and intrinsic velocity dispersion along the line of sight for three successively less stringent levels of membership discrimination.  In case (a) we consider as Fornax members only the N=176 stars surviving all four iterations of the velocity rejection algorithm; in case (b) we reinstate the six stars rejected by iterations 3 and 4; in case (c) we further add the four stars rejected by iteration 2.  With 68\% confidence intervals about the ($\hat{\langle u \rangle},\hat{\sigma_p}$) pairs in units of km s$^{-1}$, case (a) gives global values ($53.3 \pm 0.8$, $11.1 \pm 0.6$); case (b) gives ($53.0 \pm 0.9$, $12.4 \pm 0.8$); case (c) gives ($52.6 \pm 1.0$, $13.3 \pm 0.8$).  
\subsubsection{Velocity Dispersion Profile}
\begin{figure*}
  \plotone{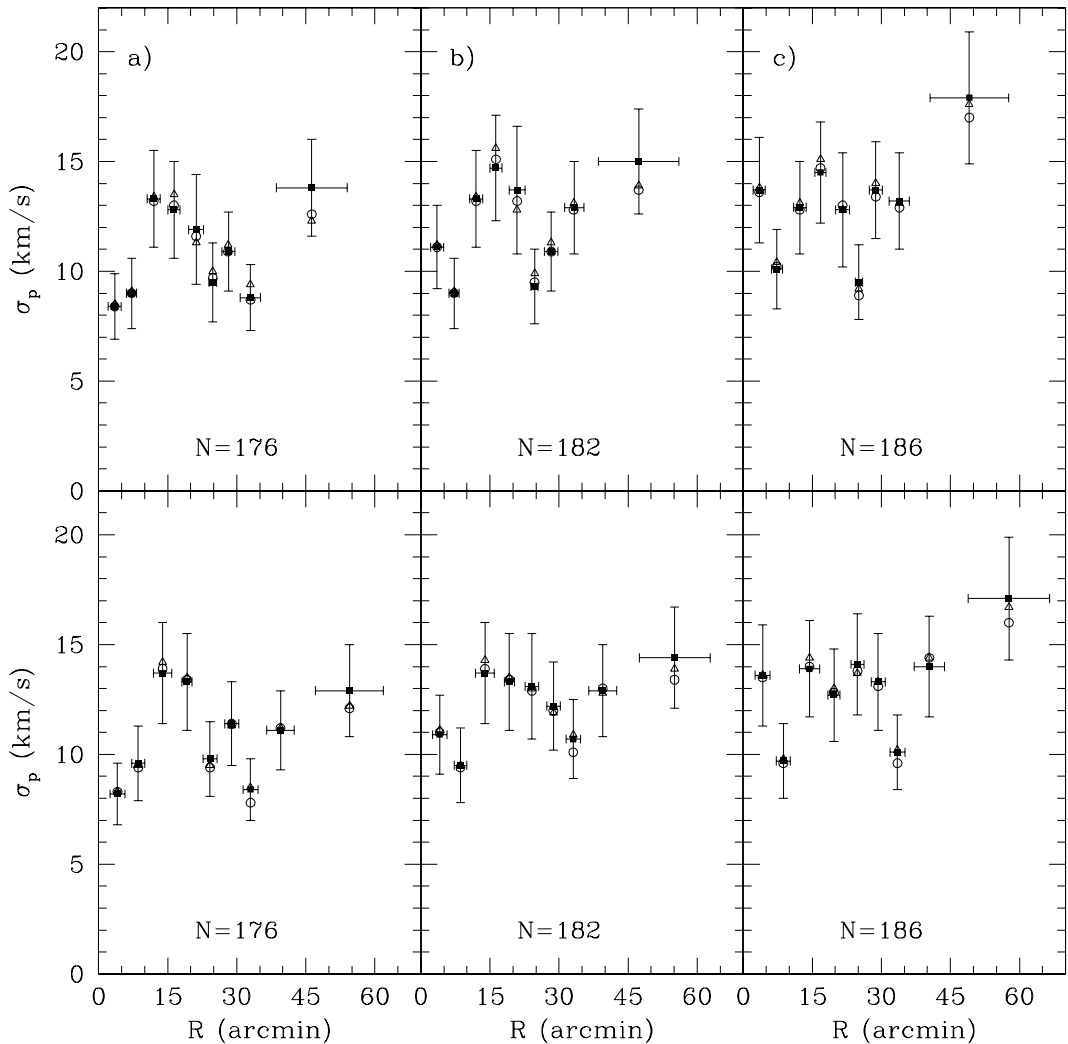}
  \caption{\label{fig:profile} \scriptsize Radial velocity dispersion as a function of angular radius for three levels of Fornax membership discrimination (see section \ref{subsec:membership}).  Filled squares and errorbars correspond to HRF radial velocity dispersion.  Open triangles and open circles indicate the GRF radial velocity dispersion calculated using, respectively, the Piatek et al. (2002) and Dinescu et al. (2004) values for Fornax's proper motion.  The plots in the top row are constructed using circular annuli, while those in the bottom row use elliptical annuli with $\epsilon \equiv 1-b/a=0.3$, semimajor axis $a=R$ and PA=41$^{\circ}$.  a) Calculated using only the 176 stars with velocities surviving all four iterations of the bi-weight rejection algorithm.  b) Calculated using the 182 stars with velocities surviving the first two rejection iterations.  c) calculated using the 186 stars with velocities surviving the first rejection algorithm.  Bins contain approximately equal numbers of stars.}
\end{figure*}

To examine the velocity dispersion as a function of radius, we divide the face of Fornax into nine annuli containing approximately equal numbers (19-21 per annulus) of member stars.  From the stars in each annulus, we estimate the intrinsic radial velocity dispersion using the maximum likelihood technique described above; here we modify the procedure, however, so that the estimated mean true velocity, $\hat{\langle u \rangle}$, of all bins is fixed at the value obtained from the global, unbinned sample.  Since Fornax has a measured ellipticity of $0.3\pm0.01$ (IH95), we estimate the profiles using elliptical as well as circular annuli.  The resulting radial velocity dispersion profile estimate, $\hat{\sigma}_p(R)$, for each of the three levels of Fornax membership rejection is shown in Figure \ref{fig:profile}.  The profiles using circular annuli display more scatter than those with elliptical annuli, although in general all are consistent with a flat profile to the limit of the sampled region.  

In addition to raising the overall dispersion, relaxing the membership criteria for the N=182 and N=186 samples emphasizes an upturn in the dispersion at the outermost annulus.  This feature persists when varying both the shape and number of annuli, and is not likely an artifact of the HRF apparent rotation signal (Figure \ref{fig:sinfit}a).  To demonstrate this last point we consider the outermost circular annulus from the N=182 velocity dispersion profile (Figure \ref{fig:profile}b).  This annulus contains 21 stars, spanning projected radii between 37 - 67 arcmin, and has, in units of km s$^{-1}$, $(\langle \hat{u}\rangle,\hat{\sigma}_p)=(53.0\pm3.4,15.0\pm2.4)$.  The eight stars to the northeast of the HRF apparent rotation axis (Section \ref{subsec:rotation}) have $(\langle \hat{u}\rangle,\hat{\sigma}_p)=(55.3\pm5.0,13.9\pm3.6)$, while the thirteen stars to the southwest have $(\langle \hat{u}\rangle,\hat{\sigma}_p)=(48.1\pm4.1,14.8\pm3.0)$.  The nine stars from the latter group that are located nearest the southwest \textit{corner} of Figure \ref{fig:photmap}b have $(\langle \hat{u}\rangle,\hat{\sigma}_p)=(47.5\pm5.4,16.1\pm3.9)$.  In each of these sub-annular regions the local velocity dispersion is equal (within statistical uncertainties) to the velocity dispersion measured for the whole annulus.  Thus the adoption of a fixed mean velocity over the entire profile has not significantly inflated the calculated dispersion even where the effects of a velocity gradient would be strongest.  Rather, the dispersion measured at large radius is dominated by localized velocity scatter.  The possible rise, or at the very least, the lack of a falloff, in the outer dispersion may place Fornax in contrast with the Draco and Ursa Minor dSphs, for which velocity dispersion profiles have recently become available (Wilkinson et al. 2004; see section \ref{subsec:draco}).   

Lin \& Dong (2005) point out that a perspective-induced HRF radial velocity gradient may produce a discrepancy between HRF and GRF radial velocity dispersion, particularly at large radii.  Overplotted without errorbars in Figure \ref{fig:profile} are the GRF radial velocity dispersion profiles measured after applying Equations \ref{eq:grf} and \ref{eq:vrel} to place the individual HRF velocities in the GRF, using either of the existing Fornax proper motion measurements.  In all annuli the GRF velocity dispersion lies well within the $1\sigma$ uncertainty region of the HRF dispersion.  We conclude that for the present  data set, the HRF radial velocity dispersion profile is a suitable surrogate for the GRF profile.  

\subsubsection{Bias in the velocity dispersion estimate?}
\label{subsubsec:bias}

By adopting Equation \ref{eq:jointGaussian} to estimate velocity dispersion we implicitly assume that the stellar velocities everywhere follow a Gaussian distribution.  This cannot strictly be correct, as tidal fields will strip high-velocity stars and internal interactions will alter the velocity distribution with time.  
If the true stellar distribution function (DF) is non-Gaussian the velocity dispersion estimate, $\hat{\sigma}_p(R)$, may deviate systematically from the true velocity dispersion profile, properly calculated as the velocity moment of the stellar DF.

To investigate the bias likely to be present in our estimate $\hat{\sigma}_p(R)$ with respect to the profile calculated from a model DF, we again perform Monte Carlo simulations.  For a given model we generate 1000 artificial  data sets, each comprising stellar radial velocities for 186 stars occupying the same sky positions as the stars in our observed  data set.  The radial velocity assigned to each star is drawn randomly from the appropriate DF, integrated at each projected radius.  For individual radial velocity uncertainties we adopt the same values as calculated for the observed sample in section \ref{subsec:uncertainties}.  We bin each artificial  data set, using circular annuli of the same radii and size as those in the observed N=186  data set.  Within each annulus, we then calculate the velocity dispersion estimate, $\hat{\sigma}_p$, and the associated confidence interval for each of the 1000 subsamples.  We sum the results to obtain a function that gives the probability of measuring a given $\hat{\sigma}_p$ within that annulus, in the case that Fornax adheres to the adopted model DF.  We then compare to the velocity dispersion calculated directly from the model DF.  

One example of a plausible non-Gaussian DF is given by King (1962, 1966; Binney \& Tremaine 1987, hereafter, BT87).  The King model has radius and velocity scale parameters $r_s$ and $v_s$, as well as a third parameter, $W_0$, that specifies the value of the central potential in units of $v_s^2$.  In our simulations we adopt King models with $W_0=3.26$ and $W_0=10.0$.  These provide reasonable single-component fits to the surface brightness and flat velocity dispersion profiles, respectively.  The latter model is also close to an isothermal sphere and so should have little distortion with respect to a Gaussian DF.  In both cases we set $r_s=13.7^{\prime}$ (IH95) and leave $v_s$ as a free parameter.

\begin{figure*}
  \plottwo{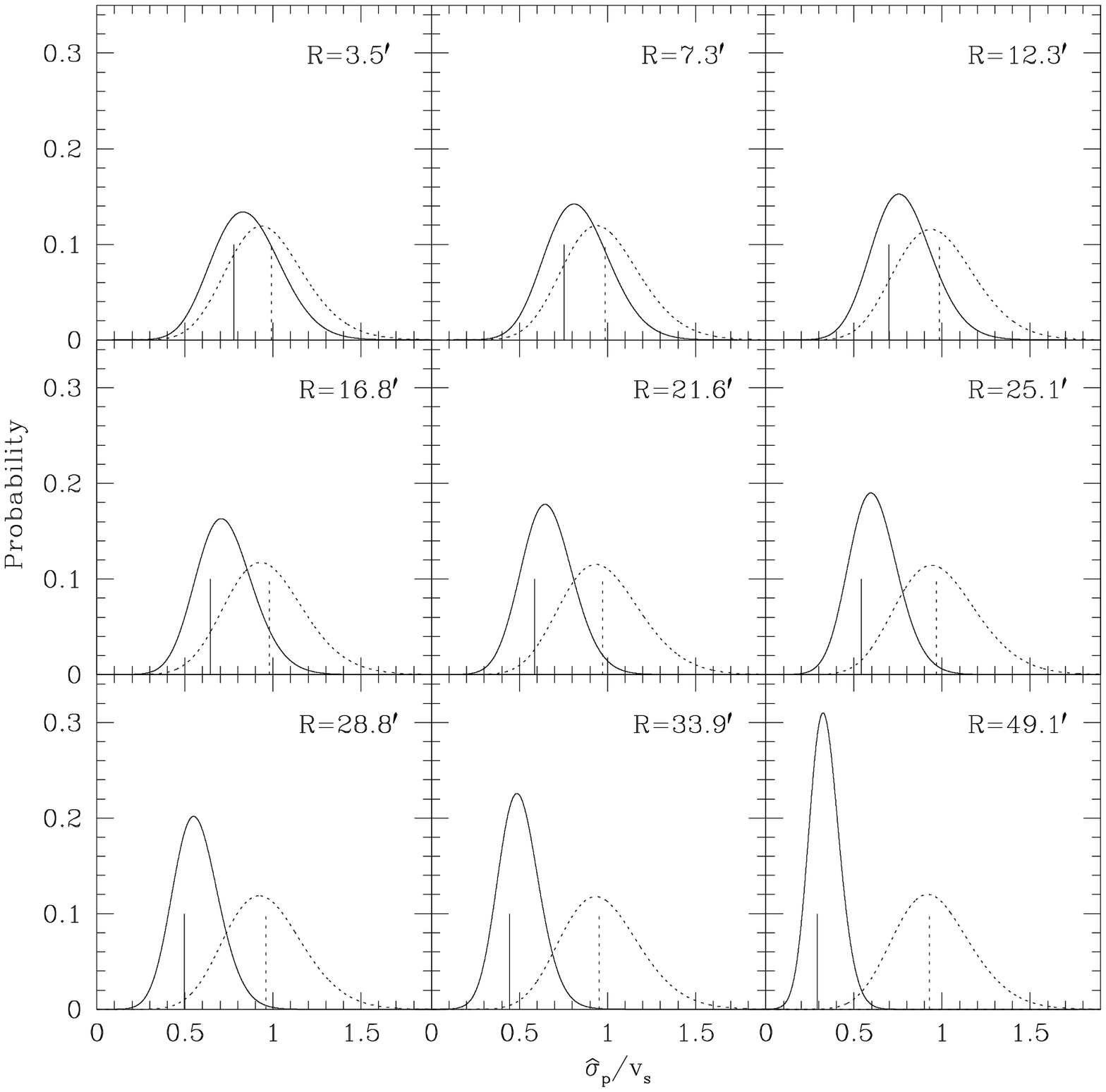}{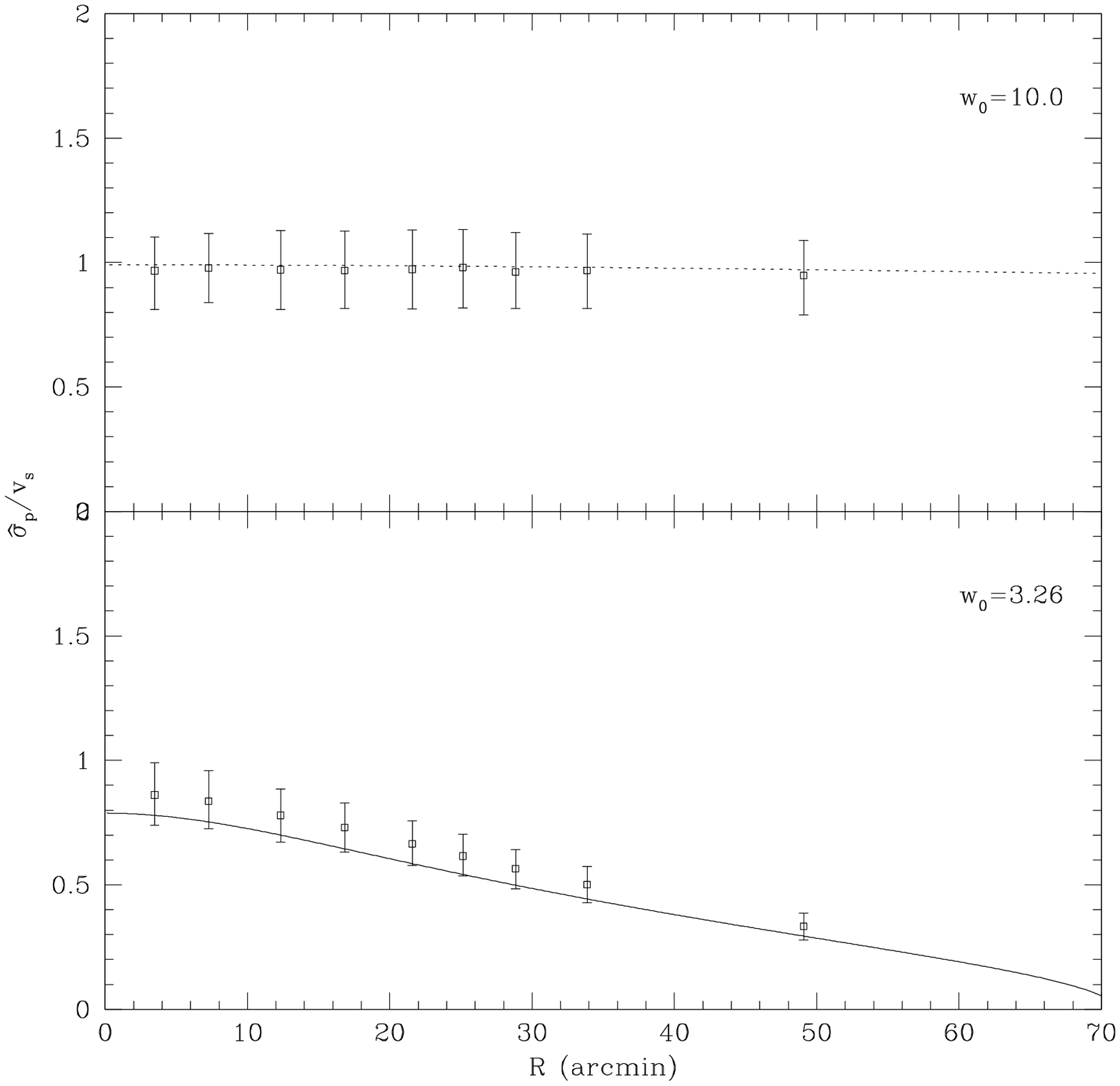}
  \caption{\label{fig:randomprofile} \scriptsize Test for bias in the velocity dispersion estimate, $\hat{\sigma}_p$, with respect to the projected velocity dispersion calculated directly from a model distribution function.  \textit{Left}: Nine panels represent the annuli (with specified angular radius) of velocity dispersion profiles calculated from simulated data.  Curves indicate the probability of measuring a velocity dispersion $\hat{\sigma}_p$ in each annulus, given 1000 simulated  data sets drawn from a King model with $W_0=3.26$ (solid) or $W_0=10.0$ (dotted).  The vertical line identifies the velocity dispersion calculated directly from the corresponding model DF.  \textit{Right}: Open squares indicate the median simulated $\hat{\sigma}_p$ in each annulus.  Errorbars enclose the $68\%$ of simulated $\hat{\sigma}_p$ values nearest the median.  Curves represent the projected velocity dispersion calculated directly from the model DF.} 
\end{figure*}

Figure \ref{fig:randomprofile} shows the resulting probability functions for $\hat{\sigma}_p$ within each annulus, and identifies the model projected velocity dispersion at the radius of the annulus.  The projected velocity dispersion for the nearly Gaussian $W_0=10.0$ model lies near the center of the simulated $\hat{\sigma}_p$ distribution in each annulus.  For $W_0=3.26$, the simulated $\hat{\sigma}_p$ values are slightly biased in favor of overestimating the model dispersion.  The discrepancy between the $W_0=3.26$ model velocity dispersion and the median simulated $\hat{\sigma}_p$ ranges from $0.04v_s$ at the outermost annulus to $0.08v_s$ at the innermost.  Since $v_s$ roughly equals the central velocity dispersion, this discrepancy amounts to $\sim 1$ km s$^{-1}$ for a dSph-like system.    

We conclude that, despite the formal distinction between the estimated velocity dispersion profile, $\hat{\sigma}_p(R)$, and the projected velocity dispersion profile calculated from a model DF, the former provides an unbiased estimate of the latter in the case of a Gaussian DF.  With respect to non-Gaussian DFs, $\hat{\sigma}_p(R)$ may introduce a bias, and one should exercise caution when comparing $\hat{\sigma}_p(R)$ to such models.  
  
\section{Analysis}
\label{sec:models}

\subsection{Equilibrium Models }
\label{subsec:king}
The classical analysis of dSph velocity  data sets has relied on application of equilibrium models falling under the purview of the core-fitting technique (Richstone \& Tremaine 1986).  Chief among these is the single-component King model, which parameterizes the stellar DF under characteristic assumptions of dynamic equilibrium, spherical symmetry, velocity isotropy, and a mass profile that is directly proportional to the luminous profile (``mass follows light'').  While they may provide a reasonable approximation of dSph cores, none of these assumptions are easily justified over the extended regions sampled by modern  data sets.  The Milky Way dSphs have measured ellipticities ranging from 0.13 to 0.56 (IH95).  The degree to which ongoing tidal interactions with the Milky Way cause departures from equilibrium is controversial and poorly constrained.  The presence of velocity anisotropy results in a well-known degeneracy with mass.  Setting aside modifications to Newtonian gravity, on no other galactic scales does mass follow light (BT87; Kormendy \& Freeman 2004).  

\begin{figure}
  \plotone{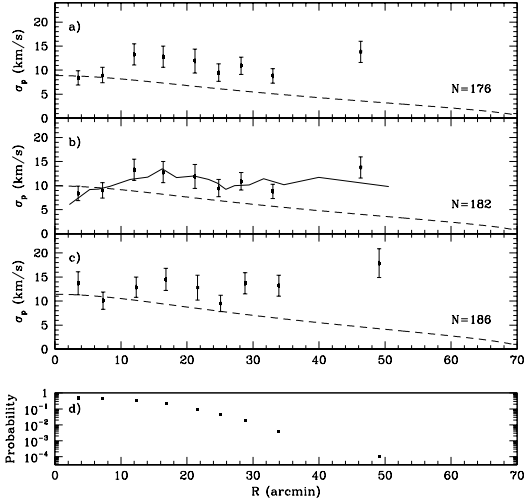}
  \caption{\label{fig:flatprob} \scriptsize (a)-(c) The projected velocity dispersion from a mass-follows-light King model ($W_0=3.26, r_s=13.7^{\prime}$; IH95) is drawn as a dashed curve over the measured $\hat{\sigma}_p(R)$ (using circular annuli), for each of three Fornax membership samples.  The solid curve in (b) is the velocity dispersion estimator of Wang et al. (2005; see section \ref{subsec:estimator}), which was calculated using this Fornax sample.  (d) Plotted for each annulus is the probability, from simulated data drawn from the mass-follows-light King DF, of measuring a velocity dispersion at least as large as the velocity dispersion of the innermost annulus.  }
\end{figure}

The velocity dispersion profile we measure for Fornax provides compelling evidence that the classical analysis is insufficient.  A single-component King model that assumes mass follows light has, adopting the photometrically-determined Fornax structural parameters of IH95, $W_0=3.26$ and $r_{s}=13.7^{\prime}$ (Figure \ref{fig:flatprob}a-c).  The artificial  data sets described in Section \ref{subsubsec:bias} indicate that, for $W_0=3.26$, the distribution of simulated $\hat{\sigma}_p$ values shifts to lower velocity dispersion toward larger angular radius (Figure \ref{fig:randomprofile}).  The median simulated $\hat{\sigma}_p$ in the outermost annulus drops to one-third the median simulated $\hat{\sigma}_p$ in the innermost annulus.  We do not see evidence of this behavior in the actual data, for which the measured velocity dispersion profile remains approximately flat at all radii.  In none of the three Fornax membership samples is the outermost measured $\hat{\sigma}_p$ less than the innermost measured $\hat{\sigma}_p$.  We use the simulated $\hat{\sigma}_p$ probability function in each annulus to calculate a negligible probability of measuring a flat velocity dispersion profile given the mass-follows-light model.  Figure \ref{fig:flatprob}d plots, for each annulus in the velocity dispersion profile, the probability (from the simulated $\hat{\sigma}_p$ distribution) that $\hat{\sigma}_p$ is at least as large as the median simulated $\hat{\sigma}_p$ in the innermost annulus.  The probability drops from $4\%$ for the annulus at R=$25.1^{\prime}$ to $2\%$ at R=$28.8^{\prime}$, then falls by an order of magnitude at each of the two remaining annuli.  We thus find a general failure of the single-component mass-follows-light King model to reproduce the flat behavior of the observed velocity disperion profile.

It is clear that \textit{at least} one of the classical assumptions is invalid in Fornax.  Perhaps the most readily discarded is mass follows light; indeed, a flat velocity dispersion profile may arise if the stars orbit inside a dark matter halo with core radius larger than that of the visible component.  We explore this scenario using two-component King models that continue to assume spherical symmetry, dynamic equilibrium, and velocity isotropy.  These models contain the additional assumption of energy equipartition in the core region, which does not readily pertain to collisionless systems such as dSphs.  We therefore adopt an approach similar to that of Pryor \& Kormendy (1990), who used two-component King models merely as tools for exploring possible dark matter distributions in the Draco and Ursa Minor dSphs\footnote{For an approach that does not assume energy coupling between the luminous and dark matter, see Wilkinson et al. (2002), who use self-consistent stellar DFs to describe stars acting as tracers in a dark plus luminous potential.  Wilkinson et al. and Pryor \& Kormendy also allow for velocity anisotropy, whereas we consider only idealized, isotropic DFs.  Radial velocity samples of the present size are only marginally able to address issues of anisotropy (Wilkinson et al. 2002), and we reserve a more comprehensive analysis for future work with larger  data sets.}.  If $E=-v^2/2-\phi$ is the total energy per unit mass, and $\phi$ is the potential per unit mass, then for the isotropic case each component $i$ has energy distribution function (King 1966; Pryor \& Kormendy 1990)
\begin{equation}
  \label{eq:king2}
  f_i(E)\propto e^{-\mu_i(E-v_s^2W_0)/v_s^2}-1.
\end{equation}
Two additional parameters join the familiar $r_s$, $v_s$, and $W_0$.  For luminous (subscript $_L$) and dark (subscript $_D$) components, $\rho_{0D}/\rho_{0L}$ and $\mu_D/\mu_L$ specify the ratios of central densities and dimensionless ``masses,'' respectively.  The $\mu_i$ have a physical interpretation when multi-component models are applied to stellar mass classes in globular clusters (e.g., Gunn \& Griffin 1979; Da Costa \& Freeman 1976).  There, for mass class $i$, $\mu_i= m_i/\overline{m}$, where $\overline{m}$ is the sum of $\rho_0^{-1}\rho_{0i}m_i$ over all mass classes.  For our purposes, $\mu_D/\mu_L$ determines the ratio of core radii, $r_{cD}/r_{cL}$, given energy equipartition in the core.  The ``core radius,'' $r_{ci}$, is defined as the radius at which the projected density of component $i$ falls to half its central value.  
 
We subject each model first to constraints set by the structural parameters derived from the \textit{single-}component King fit of IH95.  We adopt $r_{cL}=390\pm36$ pc (updated for a Fornax distance of 138 kpc) and $\Sigma_0=15.7\pm5.1$ L$_{\sun}$ pc$^{-2}$ in the V band.  The model surface brightness profile, $\Sigma(R)$, is scaled by the product $r_s\rho_0[M/L]_L^{-1}$, where $\rho_0=\rho_{0D}+\rho_{0L}$ and $[M/L]_L$ is the V-band mass-to-light ratio of the luminous component, assumed to be independent of radius.  For a given model, we assign the value of $r_s$ to be that which places $r_{cL}$ at the IH95 value.  We then assign $\rho_0[M/L]_L^{-1}$ the value that recovers the IH95 value for the central surface brightness. 

The models are next constrained by the available velocity dispersion and surface brightness profiles.  The values of $r_s$ and $\rho_0[M/L]_L^{-1}$ set the velocity scale according to $9v_s^2=4\pi Gr_s^2\rho_0$ (King 1966).  For a given value of $W_0$ and an adopted $[M/L]_L$, we determine the $(\rho_{0D}/\rho_{0L},r_{cD}/r_{cL})$ pair that provides the best fit to the IH95 surface brightness profile while having a central velocity dispersion equal to the global velocity dispersion of the Fornax sample.  Of these models we consider those with velocity dispersion profiles remaining flat over the sampled Fornax region to provide the best overall agreement with the data.  Thus we are approximating the observed velocity dispersion profiles of Figure \ref{fig:profile} as perfectly flat and ignoring any bias in $\hat{\sigma}_p(R)$ with respect to the model velocity dispersion profile.  We find that the favored models tend to have large $W_0$ values, for which such bias is expected to be minimal (Section \ref{subsubsec:bias}).

\begin{figure*}
  \plotone{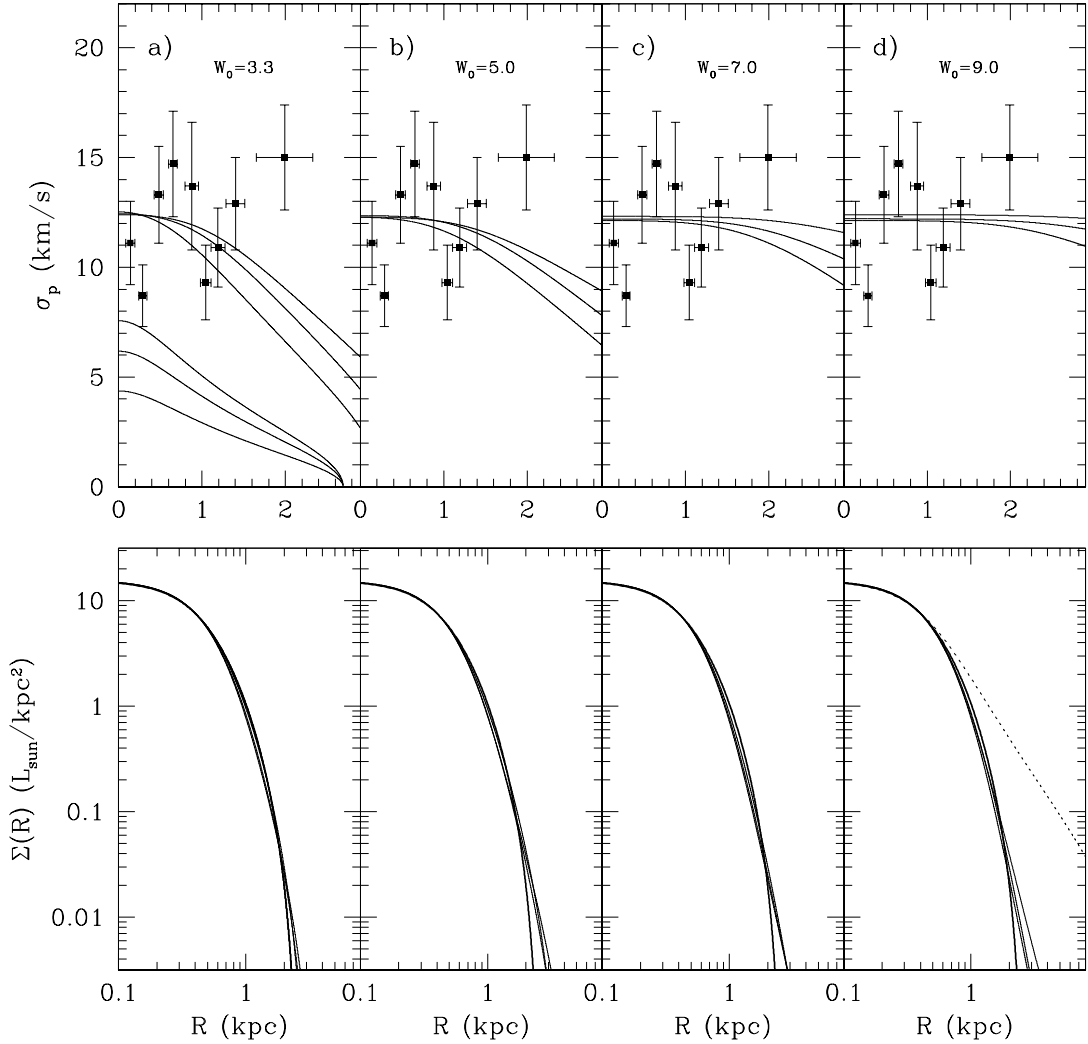}
  \caption{\label{fig:king2} \scriptsize Projected velocity dispersion (top row of panels) and surface brightness (bottom row) profiles calculated from two-component King models.  From left to right, the models have increasing values for the central potential parameter, $W_0$.  Included in the velocity plots is the observed profile from the N=182 Fornax sample.  Each set of three curves corresponds to $[M/L]_L = 1,2,3$; $\sigma_p(R)$ falls off faster for larger $[M/L]_L$.  The set of models at lower velocity dispersion in panel (a) have no dark matter, with $\rho_{0D}/\rho_{0L}=0$.  The thick curve in the surface brightness plots is a single-component model having the IH95 structural parameters.  The dotted curve in the lower panel of (d) represents a model with $W_0=9$ and $r_{cD}/r_{cL}=1.1$, and illustrates the effect of a small core radius ratio on the surface brightness profile.}
\end{figure*}

Models representing $[M/L]_L=1,2,3$ and a range of $W_0$ are summarized in Table \ref{tab:king2}.  The first column gives the number of stars considered to be members in the velocity sample.  The next six columns list the adopted $[M/L]_L$ and model parameters.  The eighth column gives $\chi^2$ per degree of freedom with respect to the velocity dispersion profile.  We do not use the $\chi^2$ test to determine a ``best-fit'' model, but merely to indicate the degree to which the considered models are consistent with a flat dispersion profile.  The final three columns list the derived quantities of interest: the central dark matter density, total mass, and overall V-band M/L.  Projected velocity dispersion and surface brightness profiles for these models are plotted in Figure \ref{fig:king2}.  For simplicity we include only those models used with the N=182 Fornax sample.  

The available Fornax velocity data place several broad constraints on the two-component models.  First, $\rho_{0D}$ must be of order $\rho_{0L}$ or larger in order to recover the observed central velocity dispersion.  This is best illustrated by the ``two''-component models with $\rho_{0D}/\rho_{0L}=0$ (Figure \ref{fig:king2}a).  These models contain no dark matter, and are therefore equivalent to the single-component, $W_0=3.3$ King model fit of IH95.  Even if $[M/L]_L=3$, models lacking a dark component underpredict the central velocity dispersion by a factor of two, and fare much worse at larger radii.  Thus it is difficult to explain the velocities of even the most central stars without invoking dark matter.  Recognizing that $\rho_{0L} \propto [M/L]_L$, the models able to reproduce the data have central dark matter densities between 0.04 - 0.10 M$_{\sun}$ pc$^{-3}$.  This is similar to the model-independent lower limit of $\rho_{0D} \geq 0.05$ M$_{\sun}$ pc$^{-3}$, derived by Pryor \& Kormendy for Draco and Ursa Minor.  

Second, the models able to reproduce the data span a surprisingly narrow range in size of the dark halo, with $2 \leq r_{cD}/r_{cL} \leq 3$.  The observed surface brightness profile helps set a $W_0$-dependent lower limit on $r_{cD}/r_{cL}$.  For models with $W_0 > 3.3$ (i.e., models for which the central potential is deeper than a single-component fit to the luminous material would suggest), the shape of the luminous density profile is sensitive to the dark matter potential.  If $r_{cD}/r_{cL}$ is sufficiently greater than unity, the dark matter density is constant over a few $r_{cL}$, and the luminous profile retains its shape even when increasing $W_0$ or $\rho_{0D}$.  To demonstrate the disruptive effect of low $r_{cD}/r_{cL}$ on the surface brightness profile, the dashed line in the bottom panel of Figure \ref{fig:king2}d is the surface brightness profile for a model with $r_{cD}/r_{cL}=1.1$.  

The apparent upper limit on $r_{cD}/r_{cL}$ may be a consequence of the assumption that the luminous and dark components are dynamically coupled.  Under Equation \ref{eq:king2}, the rate at which $\rho_{i}/\rho_0$, the density fraction of component $i$, decreases with radius is determined by the value of $\mu_iW_0$.  The density fraction for a component with large $\mu_iW_0$ declines sharply.  For large $r_{cD}/r_{cL}$, a model must have $\mu_LW_0$ sufficiently larger than $\mu_DW_0$.  However, in the limit of small $\rho_{0D}/\rho_{0L}$, $\mu_L \sim 1$ as $\mu_L/\mu_D \rightarrow \infty$.  Compared to the much less luminous Draco and Ursa Minor, Fornax favors models with smaller $\rho_{0D}/\rho_{0L}$ at smaller $W_0$.  This tends to suppress $\mu_LW_0$, thereby preventing the luminous density profile from becoming much steeper than the dark matter profile.

Finally, the flat velocity dispersion profile of Fornax favors models with large $W_0$.  As $W_0$ increases, Equation \ref{eq:king2} tends toward the distribution function of a constant velocity dispersion, isothermal sphere.  The flattening of the resulting velocity dispersion profile is evident in Figure \ref{fig:king2}.  Models with $W_0 \geq 7.0$ remain sufficiently flat over the observed region, for any $[M/L]_L \leq 3$, and so provide the best overall agreement with the data.  Models having still larger $W_0$ are not ruled out by the velocity data, although they eventually require larger values of $[M/L]_L$.  They also become unphysical, as $M \propto r$ for very large $W_0$.  

\begin{figure}
  \plotone{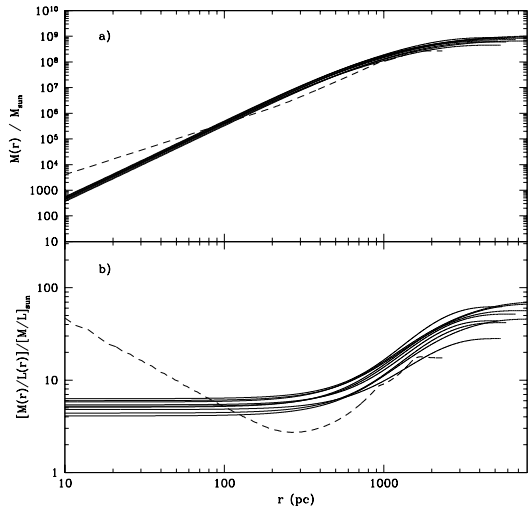}
  \caption{\label{fig:king2profiles} \scriptsize Mass and V-band $M/L$ profiles for two-component King models from Table \ref{tab:king2} with $W_0 \geq 7.0$.  (a) gives the cumulative (luminous plus dark) mass profiles.  (b) gives the cumulative mass-to-light ratio as a function of radius.  The dashed curves are nonparametric estimates (Section \ref{subsec:estimator}).}
\end{figure}

Mass and V-band $[M/L]$ profiles for the most suitable $W_0 \geq 7.0$ and $W_0=9.0$ models are shown in Figure \ref{fig:king2profiles}.  The total masses derived from these models fall in the range $4 - 18\times10^8$ M$_{\sun}$, or $3 - 7 \times10^8$ M$_{\sun}$ if we integrate the density profile only over the observed region $r < 2$ kpc.  These are one to two orders of magnitude larger than previous mass estimates based on only central velocity dispersion data (see M91 and references therein).  This dramatic increase is a consequence of the flat velocity dispersion profile.  The two-component King models suggest that if this flatness arises from stars moving isotropically inside an extended dark halo, then Fornax is very dark even over the observed region $R \leq $1$^{\circ}$, with $M/L$ perhaps 10 to 40 times larger than that of the luminous component.

%
\subsection{Non-Parametric Mass Estimation}
\label{subsec:estimator}
In Paper I we introduce a non-parametric method for estimating mass distributions from photometric and radial velocity data.  We assume spherical symmetry, velocity isotropy, and dynamic equilibrium.  We do not assume mass follows light, nor do we adopt a parametric form for the stellar DF.  Rather, we use the IH95 star count data and each velocity data point to estimate deprojected profiles for the stellar density, $f(r)$, and squared velocity dispersion, $\mu(r)$.  These relate to the underlying mass, $M(r)$, via the Jeans equations (Equation 4-55 of BT87).  In addition to shedding some of the classical assumptions, this technique offers the benefit of avoiding the problems inherent to binning a radial velocity  data set (for a King analysis that avoids binning via maximum likelihood techniques, see Oh \& Lin 1992).

Briefly, $M(r)$ is approximated as a spline of the form 
\begin{equation}
\label{eq:spline}
\hat{M}(r)=\displaystyle\sum_{i=1}^m \beta_i([r-r_{i-1}]_+)^p,
\end{equation}
in which the notation $[x]_+$ indicates the greater of $x$ or zero.  The values $\beta_1,\ldots,\beta_m$ depend on $f(r)$  and $\mu(r)$ and are estimated using the available data and by imposing general shape restrictions on $\hat{M}(r)$ (e.g., $\hat{M}(r)$ is non-decreasing and $\hat{M}(r)_{r=0}=0$).  Here we add one further shape restriction to those described in detail in Paper I.  Specifically, we require the mass density to be a decreasing function of radius (see Appendix for details).  This gives a smoother $\hat{M}(r)$, eliminating the plateau features present in the original Fornax estimation (see Figure \ref{fig:nonparametric} of the Appendix).  Simulations indicate that a strong positive bias in the mass estimate arises beyond a radius enclosing $\sim 95\%$ of the measured stars.  Inside this radius ($\sim 1.5$ kpc for the present  data set), the nonparametric technique gives $\hat{M}(r)$ accurate to within $20\%$ when operating on  data sets containing 1000 or more stellar velocities (see Figure 7 of Paper I).

We apply the nonparametric technique to the present  data set with the caveat that the uncertainty in $\hat{M}(r)$ will be at least a factor of two.  The solid line in Figure \ref{fig:flatprob}b is the nonparametric estimate of the velocity dispersion profile, $\sqrt{\hat{\mu}(r)}$, obtained using the N=182 Fornax sample and the star count data of IH95.  The corresponding $\hat{M}(r)$ is given by the dashed line in the top panel of Figure \ref{fig:king2profiles}.  The nonparametric mass estimate has a larger central value and shallower slope than the two-component King models, but displays a similar mass and behavior at radii larger than $\sim 1$ kpc.  The bottom panel of Figure \ref{fig:king2profiles} indicates that $\hat{M}(r)$ rises less steeply than the luminosity profile until approximately the Fornax core radius, where the enclosed $M/L$ reaches a minimum value of $\sim 2 [M/L]_{\sun}$.  Outside the core, dark matter dominates, reaching $M/L \sim 15 [M/L]_{\sun}$ before the estimation terminates at a radius of $1.5$ kpc.  Larger and more extended  data sets will be of great value in taking full advantage of this technique.  
%
\subsection{External Tides}
\label{subsec:tides}

Each of the mass models, as well as the nonparametric mass estimation technique we have applied to the kinematics of Fornax, assumes that Fornax is in a state close to dynamical equilibrium.  This may not be valid if the dSph stellar component is tidally heated as the galaxy orbits within the Milky Way potential.  Claims of member stars projected at distances well beyond the nominal tidal radius (as determined by a single-component King model fit to photometric data) of some dSphs have been cited as possible evidence for tidal influence (IH95; Martinez-Delgado et al. 2001; Palma et al. 2003; Mu\~{n}oz et al. 2005).  Further, the Sagittarius dSph, at a Milky Way distance of $\sim$ 16 kpc, is clearly undergoing tidal disruption (Ibata et al. 1994; Mateo et al. 1996; Ibata et al. 2001; Majewski et al. 2003), and so presents at least one case in which tides dominate.  

Various n-body simulations have addressed the degree to which the Milky Way's tidal influence on its satellites might alter their kinematics and derived masses.  Oh et al. (1995) simulate the evolution of dSphs over several perigalacticon passages and conclude that even a tidally disrupted, unbound dSph stellar population may exhibit a velocity dispersion not significantly different from its pre-disruption value.  Piatek and Pryor (1995) simulate single perigalacticon passages and add that even when a strong tidal encounter modifies the structure and internal kinematics of a dSph, the core is least affected and the central $M/L$ derived from the equilibrium assumption is virtually unchanged.  The simulations of Klessen \& Kroupa (1998) and Klessen \& Zhao (2002) show that an unbound tidal remnant projected along the line of sight may display some of the kinematic and morphological features of dSphs; however, Klessen et al. (2003) later argue that the narrow horizontal branch observed in Draco rules out the ubiquity of this scenario.  
%
%

The kinematic  data set presented here gives an opportunity to examine certain predictions of tidal disruption models.  Along a disrupting satellite's orbit, stars nearest the parent system begin to lead the satellite's center of mass as they become unbound.  Since the satellite's own gravity continues to pull on these stars in the direction opposite their motion, they lose energy in the reference frame of the parent system.  The opposite holds true for the satellite's stars farthest from the parent system: as they become unbound, they trail the satellite's center of mass and gain energy in the parent's reference frame, since the tug from the satellite is now in the same direction as their motion.  The result is elongation along the satellite's orbit, and apparent rotation of the satellite about its minor axis as observed from the parent system (Piatek $\&$ Pryor 1995; Oh et al. 1995).  Thus we might expect a disrupting satellite to display two observables: a GRF rotation signal with axis of apparent rotation perpendicular to the morphological major axis, and a proper motion parallel to the morphological major axis.  

\begin{figure}
  \plotone{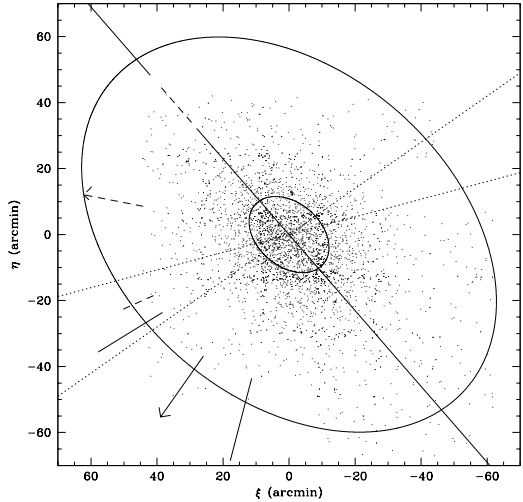}
  \caption{\label{fig:propermotionmap} \scriptsize Orientation of the apparent rotation signal and published proper motion of Fornax.   The solid line through (0,0) is the Fornax morphological major axis ($PA=41^{\circ} \pm 1^{\circ}$ (IH95)), and the dotted lines through the origin enclose the probable axes of the GRF apparent rotation signal (see section \ref{subsec:rotation}).  The receding hemisphere is to the northeast.  Arrows indicate the direction and relative magnitude of Fornax's proper motion in the Milky Way rest frame, as measured independently by Piatek et al. (2002, solid arrow), and Dinescu et al. (2004, dashed arrow).  The associated proper motion uncertainties are mapped conservatively, encompassing all directions allowed by the uncertainties quoted for the two components of proper motion.  }
\end{figure}

For Fornax, the GRF rotation signal detected using the Piatek et al. proper motion (Section \ref{subsec:rotation}) has a rotation axis oriented nearly perpendicular to the galaxy's morphological major axis (Figure \ref{fig:propermotionmap}).  The arrows in Figure \ref{fig:propermotionmap} show the directions and relative magnitudes of the Piatek et al. and Dinescu et al. proper motion measurements of Fornax.  The Piatek et al. proper motion vector is clearly not aligned with the major axis, contrary to predictions from the models of tidal interaction.  In contrast, the Dinescu et al. proper motion is more nearly parallel to the major axis.  However, the $\sim 1.2$ km s$^{-1}$ GRF rotation implied by the Dinescu et al. proper motion is too slow to have statistical significance.  Thus the two observables indicative of tidal disruption are not simultaneously present in the existing kinematic data.  
%

\begin{figure}
  \plotone{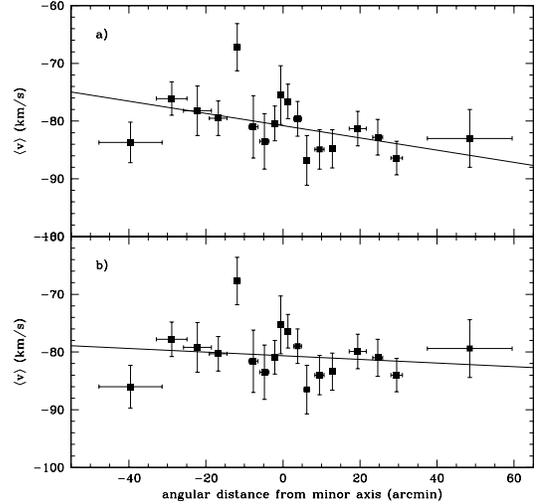}
  \caption{\label{fig:grfgradient} \scriptsize GRF mean radial velocity along Fornax's morphological major axis.  GRF velocities are calculated using Fornax proper motion measurements by Piatek et al. (2002; panel {a}) and Dinescu et al. (2004; panel {b}).  Solid lines represent the best-fitting radial velocity gradient, assuming cylindrical solid-body rotation.}
\end{figure}

If we ignore the orientation of rotation and proper motion and instead simply examine the major-axis velocity trend, we find no evidence for a tidally-induced velocity gradient along this axis.  For a star with GRF radial velocity $v$, let $D$ be the angular distance between the star and Fornax's minor axis (i.e., distance \textit{along} the major axis), and let $v_{sys}$ be the bulk GRF radial velocity of Fornax.  We model $dv/dD$, the GRF radial velocity gradient along the major axis, according to $v(D)=v_{sys}+dv/dD$.  This assumes that any apparent rotation resembles that of a cylindrical solid body.  We then use the unbinned GRF velocity data to solve for $dv/dD$ via linear regression.  Using the Piatek et al. proper motion, we find $dv/dD\sim 0.1$ km s$^{-1}$ arcmin$^{-1}$ (3 km s$^{-1}$ kpc$^{-1}$), and the Dinescu et al. proper motion gives a shallower $dv/dD\sim 0.03$ km s$^{-1}$ arcmin$^{-1}$ (0.8 km s$^{-1}$ kpc$^{-1}$).  These gradients are drawn over plots of the mean velocity along the major axis in Figure \ref{fig:grfgradient}.  The unbinned data display significant scatter about either gradient, with each fit having $\chi^2/$dof $ \sim 30$.  This contrasts with the predictions of tidal disruption models (see, Figure 7 of Piatek \& Pryor 1995; Figure 5 of Klessen \& Zhao, 2002) in which tides produce ordered, monotonic, and typically steeper gradients.  The dispersion about either of the best-fit gradients is $\sim$ 12 km s$^{-1}$, identical to the overall sample velocity dispersion.  This indicates that the Fornax kinematics are dominated by random motions, rather than the ordered, streaming motions indicative of tidal disruption.  
 
It should be noted that tides are not the only mechanism by which a dSph might be altered from a state of dynamic equilibrium.  During a search for extratidal structure, Coleman et al. (2004; 2005) discovered two lobes along the Fornax minor axis and aligned with two shell-like features.  They interpret these as signs of a recent merger event, rather than tidal tails.  If a recent merger is confirmed, then there are localized regions within Fornax that have not had time to virialize, making irrelevant the concept of a virial tidal radius.  To what extent such mergers may be pervasive in the dSph population remains highly uncertain, although Kleyna et al. (2003; 2004) have detected kinematically distinct substructure in Ursa Minor and Sextans.  In addition, Tolstoy et al. (2004) have found evidence for two populations of ancient stars with differing metallicity \textit{and} velocity distributions within the Sculptor dSph.  It is clear that dSphs have more complicated histories than once thought, and many more stellar spectra are required to identify distinct components.

\section{Discussion}
\label{sec:discussion}

\subsection{Astrophysical Velocity Variability}
\label{subsec:binarism}
We have assumed thus far that the measured radial velocities result exclusively from the underlying gravitational potential and kinematic status of Fornax.  This is not necessarily true.  Binary orbital motion may add a random component to any single-epoch velocity measurement, as may bulk stellar atmospheric motions.  

Including the M91 data, the Fornax  data set now contains multi-epoch velocity measurements for 20 stars.  Having no more than five, and in most cases only two, distinct measurements for any one of these stars, we cannot deduce binary parameters.  Instead we identify binary candidates as those stars exhibiting velocity variability exceeding that which we would expect from the formal measurement errors.  For each multiply-observed star, we calculate the $\chi^2_{obs}$ obtained from the velocity measurements and their formal errors (see Tables 3 and 4), as well as the probability, $p(\chi^2_{obs})$, that $\chi^2\geq\chi^2_{obs}$  given Gaussian random measurement errors.  These probabilities should follow a uniform distribution between 0.0 and 1.0 if the stars are non-variable.  For our sample of 20 multiply-observed stars, this would predict roughly two stars having $p(\chi^2_{obs})$ falling within each probability range 0.0-0.1, 0.1-0.2, ..., 0.9-1.0, but only 0.02 stars having $p(\chi^2_{obs})\leq0.001$, which is the probability threshold suggested by Olszewski et al. (1996, hereafter, OPA96) as indicative of binarity.  While we do indeed find between 1-3 stars per each tenth in probability between 0.0 and 1.0, there are two stars---F-M20 and F2-9---for which $p(\chi^2_{obs})\leq0.001$.  The excess over the expected number of stars at very low $p(\chi^2_{obs})$ suggests that at least these two stars are exhibiting true velocity variability and are therefore binary candidates.  This implies a binary ``dicovery fraction'' of 0.1 for the present Fornax sample.

The actual Fornax binary frequency depends not only on the fraction of observed stars we can identify as binaries, but also on the efficiency with which we can identify binaries among our subset of multiply-observed stars.  OPA96 perform simulations over an expanse of binary orbital parameters in order to examine the discovery efficiency for a sample of 118 stars with multiple velocity measurements in the Draco and Ursa Minor dSphs.  We do not attempt to replicate their procedure regarding the present Fornax sample, primarily because OPA96 have superior statistics from multiple measurements.  We wish to emphasize, however, the important conclusion from OPA96 that the presence of binary stars in a dSph radial velocity sample ultimately has little effect on the derived velocity dispersion.  The error in the velocity dispersion due to sampling uncertainty outweighs the error introduced by binaries.  Simulations show that this result is a general feature of dSph-like velocity samples given measurement deviations similar to those in the Draco-UMi sample (OPA96 Table 7; see also Hargreaves et al. 1996, who reach a similar conclusion from binary simulations).  

A binary discovery fraction of 0.1 suggests the presence of at least 20 unidentified binaries in our Fornax velocity sample of $\sim$ 200 stars.  We make a crude attempt to examine the effect of unidentified binaries on the derived Fornax properties by observing the effects of removing the two known binaries from various subsamples of the velocity  data set.  First, if we calculate the velocity dispersion for the 20 stars with multiple velocity measurements, we obtain $\hat{\sigma}_{multiples}=15.8\pm2.7$ km s$^{-1}$.  If we remove the probable binaries F-M20 and F2-9 we calculate a slightly \textit{larger} value, $\hat{\sigma}_{multiples}=16.4\pm3.0$ km s$^{-1}$.  Considering the velocity dispersion profile, F-M20 is in the innermost bin, and F2-9 is in the third bin.  If we re-calculate the velocity dispersions in these bins after the removal of the candidate binary (both bins originally contain 20 stars, so based on our binary discovery frequency, we expect at least 1 undetected binary to remain in each bin), we find that removal of F-M20 causes the dispersion estimate in the innermost bin to \textit{rise} from $11.3\pm2.0$ km s$^{-1}$ to $11.5\pm2.1$ km s$^{-1}$, and the removal of F2-9 causes the dispersion estimate in the third bin to rise from $13.4\pm2.4$ km s$^{-1}$ to $13.7\pm2.6$ km s$^{-1}$.  In both cases, inclusion of the probable binary has negligible impact on the measured velocity dispersion.  We can draw no strong conclusion from these tests, as the number of detected binaries is small and the discovery efficiency is unknown.  Nevertheless, we find nothing to refute the conclusion of OPA96 that binary stars negligibly inflate the measured velocity dispersion.

A second possible source of radial velocity noise may come in the form of bulk motions in the atmospheres of the observed stars (Gunn \& Griffin 1979).  Pryor et al. (1988) find this velocity ``jitter'' to be as large as 4-8 km s$^{-1}$ in globular cluster red giants, although the effect appears to fall off rapidly in stars more than 0.5 mag dimmer than the tip of the giant branch.  If we define the tip of the Fornax RGB to have I $\sim$ 16.7 (Figure \ref{fig:cmdtrue}), then the region susceptible to velocity jitter includes 47 stars from the N=186 Fornax sample (all but one of these stars were originally selected for spectroscopic observation prior to the observation of the photometric  data set presented in section \ref{sec:velocities}).  Of these 47, twelve have repeat velocity measurements that we may examine for variability.  Only one of these, F-M20---identified in the previous section as a binary candidate---has $p(\chi^2_{obs})\leq0.01$.  While the velocity variability of the dimmer binary candidate F2-9 is unlikely to be due to atmospheric jitter, the variability of F-M20 may be due in part to atmospheric motion.  The distribution of $p(\chi^2_{obs})$ is otherwise uniform.  The velocity dispersion calculated from the 47 brightest giants is $13.4 \pm 1.5$ km s$^{-1}$.  The velocity dispersion calculated from the remaining 139 stars from the N=186 sample is $12.9 \pm 0.8$ km s$^{-1}$.  We conclude that atmospheric jitter does not have a significant impact on the Fornax radial velocities.

\subsection{Comparison with Draco and Ursa Minor}
\label{subsec:draco}

Velocity dispersion profiles extending to the limits of the stellar distributions are now available for the dSphs Fornax, Ursa Minor, and Draco (Kleyna et al. 2002; Wilkinson et al. 2004, hereafter, WK04;, Mu\~{n}oz et al. 2005).  The behavior of the dispersion profiles at large radius is of great interest, capable not only of distinguishing between kinematic models, but also testing assumptions on which those models are based and searching for tidal influence on the kinematics.  Although the profiles are generally flat, WK04 find sharp \textit{decreases} in the velocity disperions of both Draco and Ursa Minor at the outermost point of both profiles.  The observed drops are too sudden to be explained by isotropic King or Plummer profiles.  although an abrupt change in the velocity anisotropy might explain the drop observed in Draco (see Mashchenko et al. 2005), WK04 argue that anisotropy cannot by itself provide a plausible explanation for the more dramatic drop they witness in Ursa Minor.  If the sharply declining dispersions in Draco and Ursa Minor are real features, they indicate an absence of tidal heating at the large radii at which they occur (Read et al. 2005).  Read et al. argue that dark matter halos of up to $10^{9} - 10^{10}$ M$_{\sun}$ are necessary to prevent tidal heating of the stars in these dSphs.  This would indicate similarity between Draco, Ursa Minor, and Fornax as described by the isotropic, two-component King models of section \ref{subsec:king}.  It should be noted that Mu\~{n}oz et al. (2005) have recently reanalyzed the data of Wilkinson et al. and conclude that the presence of such a drop depends largely on the binning scheme and membership criteria employed.  

Regarding Fornax, we find no evidence for a decrease in the velocity dispersion at large radius for any binning scheme or reasonable membership criteria.  Instead, we see some evidence for a mild increase at the outermost data point, particularly as we allow stars from the wings of the observed velocity distribution into the Fornax membership.  If this rise were attributable to tidal effects it would be puzzling that Fornax is susceptible to such external influence while Draco and Ursa Minor are not.  Fornax has considerably larger luminous mass ($L_V=1.5\times 10^7L_{\sun}$; Mateo 1998) and lies at a greater distance from the Milky Way (the two proper motion studies cited in this work estimate that the current distance of $\sim$ 140 kpc is near perigalacticon) than either Draco ($L_V=2.6 \times 10^5L_{\sun}$, $D=82 \pm 6$ kpc) or Ursa Minor ($L_V=2.9 \times 10^5L_{\sun}$, $D=66 \pm 3$ kpc).  The degree to which the Milky Way influences the kinematics of its satellites remains an open and intriguing question.  The remaining Galactic dSphs---Sculptor ($L_V=2.2 \times 10^6 L_{\sun}$, $D=79\pm4$ kpc), Sextans ($L=5.0 \times 10^5 L_{\sun}$, $D=86\pm4$ kpc), Leo I ($L=4.8 \times 10^6 L_{\sun}$, $D=250\pm30$ kpc), Leo II($L=5.8 \times 10^5 L_{\sun}$, $D=205\pm12$ kpc), Carina ($L=4.3 \times 10^5 L_{\sun}$, $D=101\pm5$ kpc), and the recently discovered Ursa Major dSph ($L_V \sim 4 \times 10^4 L_{\sun}$, $D \sim 100$ kpc; Willman et al. 2005)---occupy a large region of parameter space.  High-quality velocity  data sets are necessary to determine any correlation between mass, orbital parameters, and the behavior of the dispersion profile at large radius.

\subsection{Summary and Conclusions}       
\label{sec:summary}

We have presented new radial velocity measurements for 156 (+9 with uncertain membership status) stars belonging to the Fornax dSph.  This increases the total number of Fornax stars with published velocities to 176 (+10).  In order to test for rotation we have used existing Fornax proper motion measurements to place the heliocentric velocities in the Fornax rest frame.  Adoption of the Piatek et al. (2002) proper motion results in a (marginally) statistically significant GRF rotation signal of $\sim 2.5$ km s$^{-1}$ about an axis at $112^{\circ}\pm8^{\circ}$, near Fornax's minor axis.  Adoption of the Dinescu et al. (2004) proper motion results in no statistically significant GRF rotation signal.  Despite a favorable orientation with respect to the minor axis, the rotation signal stemming from the Piatek et al. proper motion is difficult to attribute to tidal influence, as the proper motion vector is perpendicular to Fornax's morphological major axis.  Thus the two predictions from tidal disruption models---apparent rotation about the minor axis and elongation along the satellite's orbit---are not simultaneously evident in the present data.  By examining localized velocity dispersions and the velocity gradient along the major axis, we have demonstrated that the stellar kinematics of Fornax are dominated by random, rather than bulk rotational or streaming tidal motions.

The Fornax radial velocity dispersion profile is generally flat.  We have demonstrated the inability of single-component King models to account for the observed velocity dispersion profile.  We have applied isotropic, two-component King models consistent with the observed Fornax surface brightness profile, and found that models having similar central densities for dark and luminous matter are able to reproduce the flat observed velocity dispersion profile if the dark matter halo has a core of at least twice the size of the luminous material.  Two-component models favored by the data have masses in the range $M \sim 10^{8} - 10^{9}$ M$_{\sun}$.  This would indicate a similarity between Fornax and the dSphs Draco and Ursa Minor, if external tides indeed do not affect the stellar kinematics in these systems.  In this case, dSphs are even more massive and dark-matter dominated than previously thought, which may help ameliorate the ``accounting problem'' faced by cold dark matter models (Klypin et al. 1999; Moore et al. 1999; Stoehr et al. 2002).  

In any case, these results add to the mounting evidence that we must turn to more sophisticated analytical tools in order to explain the motions of dSph stars, particularly at large radii.  We have discussed one such tool---a nonparametric statistical smoothing technique for estimating spherical masses directly from radial velocity data.  An application of this method to the present Fornax sample yields a model-independent estimate of the mass profile.  The result is consistent with the large Fornax masses suggested by the two-component models.  This will become a powerful tool as dSph  data sets continue to grow.  We direct the interested reader to Paper I, in which this method is introduced in formal detail, and to the appendix to this work.  

We thank the staff at Las Campanas Observatory for support during these observing runs.  We are grateful to Slawomir Piatek and Tad Pryor for valuable comments and discussion.  We also thank the referee, D.N.C. Lin, for comments and criticisms that improved this work.  We thank Ewa Lokas for calling to our attention the fact that in the preprint version of this article, repeat observations of three stars were listed in Table 1 with unique stellar IDs.  This has been corrected in the present version, and calculated quantities have been updated (results are not affected significantly).  MM and MGW are supported in part by NSF grants AST-0507453, AST-0206081, and previously by NSF grant 0098661.  EO is supported in part by NSF grants AST-0098518, AST-0205790, and AST0507453.  This work also received funding from the Horace H. Rackham School of Graduate Studies at the University of Michigan, Ann Arbor (Rackham Interdisciplinary Collaboration Research Grant).

\begin{figure}
  \plotone{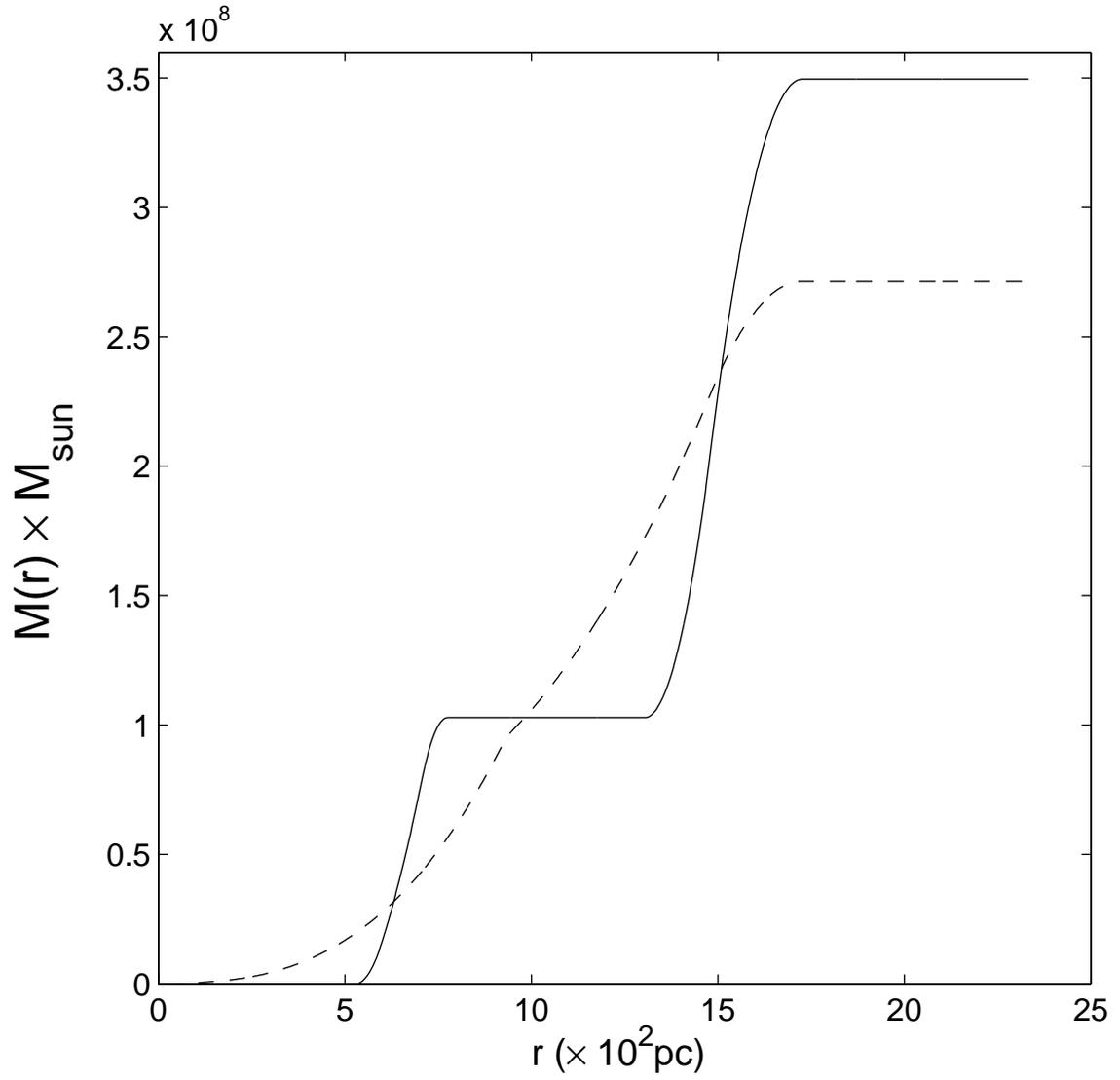}
  \caption{\label{fig:nonparametric} \scriptsize Nonparametric estimation of the Fornax mass.  The solid curve is the original result published in Wang et al. 2005.  The dashed curve is the estimate produced using the same data, under the additional constraint that the mass density is a non-increasing function of radius.}
\end{figure}

\begin{appendix}
  Wang et. al. (2005) modelled $M$ by a quadratic spline,
\begin{equation}
  \label{eq:qdspln}
  \hat{M}(r)=\sum_{i=1}^m \beta_i(r-r_{i-1})_{+}^{2}
\end{equation}
and reduced the estimation problem to a quadratic programming problem in which a quadratic function $Q(\beta_1,\cdots,\beta_m)$ is to be minimized subject to the constraints
\begin{equation}
  \label{eq:cnstrnts1}
  \sum_{i=1}^m \beta_i = 0 =\sum_{i=1}^m \beta_i(r_m-r_{i-1})
\end{equation}
and
\begin{equation}
  \label{eq:cnstrnts2}
  \sum_{i=1}^j \beta_i(r_j-r_{i-1}) \ge 0
\end{equation}
for $j = 1,\cdots,m-1$. These constraints are necessary for a quadratic spline to be non-negative, non-decreasing, and bounded.  Denoting the solution to the quadratic programming problem by $\hat\beta_1,\cdots,\hat\beta_m$, the resulting estimator, given by Equation \ref{eq:qdspln},
tracked the gross features of (the true) $M$ quite well in simulations, but it often has flat stretches, implying regions of no estimated mass (Figure \ref{fig:nonparametric}, solid line). The latter feature can be eliminated (as in the dashed line in Figure \ref{fig:nonparametric}) by supposing that the mass density $\rho(r)$ is non-decreasing in $r$.  This assumption leads to the further constraints
\begin{equation}
  \label{eq:cnstrnts3}
  \sum_{i=1}^j \beta_i[2r_{i-1}-r_k] \le 0 
\end{equation}
for $k = j-1$ and $k = j$ for $j = 1,\cdots,m$.  Estimation of $M$ with the additional assumption that $\rho$ is non-increasing then proceeds as in Paper I: the estimated $M$ is given by (\ref{eq:qdspln}), where now $\hat\beta_1,\cdots,\hat\beta_m$ denote the solution to the quadratic programming problem with the constraints (\ref{eq:cnstrnts3}) in addition to (\ref{eq:cnstrnts1}) and (\ref{eq:cnstrnts2})

It is also possible to estimate the density, since
$$
\rho(r) = {1\over 4\pi r^2}M'(r),
$$
where the prime denotes derivative.  For $\hat{M}$ this becomes
\begin{equation}
  \label{eq:estrho}
  \hat\rho(r) = \sum_{i=1}^m 2\hat\beta_i{(r-r_{i-1})_{+}\over 4\pi r^2}
\end{equation}
Unfortunately, (\ref{eq:estrho}) becomes infinite as $r \rightarrow 0$, a feature that is forced by the use of quadratic splines.  This is not a fundamental problem as the total mass remains finite, nor is it unusual within the realm of widely used dynamical models  (see $\rho(r)$ for the NFW model (Navarro et al. 1997),
or the classical isothermal sphere).

To see how the additional constraints (\ref{eq:cnstrnts3}) arise, first observe that $\rho$ is non-increasing if and only if
$$
{d\over dr}\big[{1\over r^2}M(r)\big] \le 0.
$$
For $M$ of the form (\ref{eq:qdspln}),
$$
{d\over dr}\big[{1\over r^2}M(r)\big] = 2\sum_{i=1}^j \beta_i \big[{1\over r^2} - 2{r-r_{i-1}\over r^3}\big]
$$
in the interval $r_{j-1} \le r < r_j$ for $j = 1,\cdots,m$.  So,
$$
r^3{d\over dr}\big[{1\over r^2}M(r)\big] = 2\sum_{i=1}^j \beta_i[2r_{i-1}-r]
$$
in the interval $r_{j-1} \le r < r_j$ for $j = 1,\cdots,m$.  Viewed as a function on the interval $[0,r_m]$, the latter is a discontinuous, piecewise linear function.  So, it will be non-positive everywhere if and only if it is non-positive at the endpoints of each interval $r_{j-1} \le r < r_j$.

\end{appendix}
\clearpage
\LongTables
\renewcommand{\arraystretch}{0.6}
\begin{deluxetable*}{lcccccccccl}
  \tablenum{1}
  \tabletypesize{\scriptsize}
  \tablewidth{0pc}
  \tablecaption{\scriptsize Heliocentric Radial velocity results for Fornax Target Stars}
  \label{tab:results}
  \tablehead{\colhead{Star} & \colhead{$\alpha_{2000}$} & \colhead{$\delta_{2000}$} & \colhead{HJD} & \colhead{UT Date} & \colhead{R} & \colhead{PA} & \colhead{I} & \colhead{V-I} & \colhead{$v$} & \colhead{Member}\\
    \colhead{}&\colhead{}&\colhead{}&\colhead{(-2400000.0)}&\colhead{of observation}&\colhead{(arcmin)}&\colhead{(degrees)}&\colhead{}&\colhead{}&\colhead{(km s$^{-1}$)}& \colhead{}
  }
  \startdata
F1-1     & 02:38:51.5& -34:35:25.5& 48955.6& 29 Nov 1992&   14.5& 120.2&  16.18&   2.21&    $55.8\pm2.7$& Y\\
     & & & 52623.5& 15 Dec 2002& & & & &    $64.5\pm2.0$& \\
F1-2     & 02:38:52.3& -34:44:55.3& 48955.7& 29 Nov 1992&   20.8& 143.7&  16.41&   1.64&    $42.9\pm2.6$& Y\\
     & & & 48961.6& 05 Dec 1992& & & & &    $45.2\pm2.2$& \\
F1-3     & 02:38:59.9& -34:45:26.9& 48955.7& 29 Nov 1992&   20.4& 148.1&  16.54&   1.72&    $49.6\pm2.3$& Y\\
F1-4     & 02:39:53.3& -34:46:02.6& 48955.8& 29 Nov 1992&   17.9& 180.7&  16.41&   1.83&    $44.7\pm2.6$& Y\\
     & & & 52621.6& 13 Dec 2002& & & & &    $43.5\pm1.7$& \\
     & & & 52622.6& 14 Dec 2002& & & & &    $43.5\pm1.7$& \\
F1-5     & 02:38:03.0& -34:16:39.1& 48956.6& 30 Nov 1992&   25.3&  63.1&  16.74&   1.55&    $45.5\pm2.7$& Y\\
F1-6     & 02:39:14.4& -34:22:42.0& 48957.7& 01 Dec 1992&    9.5&  55.2&  16.84&   1.53&    $73.9\pm2.6$& Y\\
F1-7     & 02:38:23.7& -34:16:38.5& 48957.7& 01 Dec 1992&   21.6&  57.9&  17.23&   1.45&    $61.3\pm2.8$& Y\\
F1-8     & 02:38:26.5& -34:15:01.6& 48957.7& 01 Dec 1992&   22.0&  53.6&  17.17&   1.44&    $48.5\pm2.8$& Y\\
F1-9     & 02:39:28.1& -34:26:02.2& 48957.8& 01 Dec 1992&    5.4&  67.1&  16.76&   1.63&    $42.7\pm2.8$& Y\\
F1-10     & 02:39:52.3& -34:18:25.5& 48958.6& 02 Dec 1992&    9.7&   0.0&  17.03&   1.48&    $51.3\pm2.7$& Y\\
F1-11     & 02:40:07.2& -34:19:18.1& 48958.6& 02 Dec 1992&    9.4& 340.8&  16.86&   1.57&    $55.1\pm2.6$& Y\\
F1-12     & 02:40:15.3& -34:18:09.2& 48958.6& 02 Dec 1992&   11.1& 334.6&  16.72&   1.58&    $41.5\pm2.8$& Y\\
F1-13         & 02:37:55.8& -34:47:16.0& 48958.7& 02 Dec 1992&   30.7& 128.7&  16.78&   1.63&    $33.5\pm2.2$& Y\\
F1-14     & 02:40:14.1& -34:22:57.9& 48958.7& 02 Dec 1992&    6.9& 319.1&  17.05&   1.46&    $43.7\pm2.3$& Y\\
F1-15       & 02:38:09.0& -34:51:07.1& 48958.8& 02 Dec 1992&   31.3& 137.4&  16.92&   1.58&    $55.0\pm2.7$& Y\\
F1-16     & 02:39:11.1& -34:39:09.0& 48959.6& 03 Dec 1992&   13.9& 142.4&  16.74&   1.62&    $60.6\pm2.7$& Y\\
F1-17     & 02:39:36.8& -34:45:29.7& 48959.7& 03 Dec 1992&   17.6& 169.6&  16.86&   1.54&    $67.7\pm2.6$& Y\\
F1-18     & 02:39:40.0& -34:43:00.9& 48959.7& 03 Dec 1992&   15.1& 170.4&  16.84&   1.50&    $28.5\pm2.8$& Y\\
F1-19     & 02:39:42.7& -34:43:33.4& 48959.7& 03 Dec 1992&   15.5& 172.7&  16.84&   1.62&    $35.5\pm2.8$& Y\\
F1-20     & 02:38:43.5& -34:32:05.5& 48959.7& 03 Dec 1992&   14.7& 105.6&  16.82&   1.47&    $63.0\pm2.8$& Y\\
F1-21     & 02:38:57.9& -34:28:26.7& 48959.7& 03 Dec 1992&   11.2&  91.6&  16.91&   1.43&    $38.6\pm2.6$& Y\\
F1-22     & 02:38:53.6& -34:33:04.8& 48959.8& 03 Dec 1992&   13.1& 112.3&  16.71&   1.63&    $39.8\pm2.6$& Y\\
F1-23     & 02:38:53.8& -34:30:06.8& 48959.8& 03 Dec 1992&   12.2&  99.3&  16.80&   1.59&    $71.1\pm2.7$& Y\\
F1-24     & 02:39:00.7& -34:33:17.0& 48960.6& 04 Dec 1992&   11.8& 115.9&  16.75&   1.44&    $54.9\pm2.5$& Y\\
F1-25     & 02:38:49.3& -34:24:05.2& 48960.6& 04 Dec 1992&   13.6&  72.7&  16.89&   1.53&    $50.5\pm2.8$& Y\\
F1-26     & 02:39:08.0& -34:19:11.8& 48960.6& 04 Dec 1992&   12.8&  45.6&  17.06&   1.43&    $67.5\pm2.9$& Y\\
F1-27     & 02:39:15.8& -34:17:43.2& 48960.6& 04 Dec 1992&   12.9&  35.9&  17.18&   1.41&    $30.0\pm2.9$& Y\\
F1-28     & 02:39:06.1& -34:37:29.7& 48960.7& 04 Dec 1992&   13.3& 134.5&  16.90&   1.45&    $45.2\pm2.7$& Y\\
F-M18     & 02:39:31.5& -34:31:50.9& 48961.7& 05 Dec 1992&    5.7& 130.8&  16.15&   1.82&    $55.0\pm2.8$& Y\\
F1-30     & 02:39:30.6& -34:24:07.8& 48961.8& 05 Dec 1992&    6.0&  48.1&  16.32&   1.85&    $50.2\pm2.8$& Y\\
F-M26     & 02:39:40.1& -34:34:02.3& 48963.7& 07 Dec 1992&    6.4& 156.9&  16.36&   1.79&    $41.2\pm2.3$& Y\\
       & & & 49341.6& 20 Dec 1993& & & & &    $43.8\pm2.0$& \\
       & & & 49649.6& 23 Oct 1994& & & & &    $40.3\pm2.5$& \\
F-M15       & 02:39:54.0& -34:34:24.3& 49333.6& 12 Dec 1993&    6.3& 183.1&  15.90&   1.96&    $61.7\pm1.8$& Y\\
F-M4        & 02:40:01.8& -34:27:48.1& 49333.6& 12 Dec 1993&    2.0& 280.1&  16.82&   1.63&    $52.3\pm2.0$& Y\\
        & & & 49648.6& 22 Oct 1994& & & & &    $53.4\pm2.6$& \\
F18-1       & 02:41:17.6& -34:13:07.2& 49333.6& 12 Dec 1993&   23.1& 310.4&  16.98&   1.50&    $50.0\pm2.0$& Y\\
F15-1       & 02:42:10.0& -34:18:16.5& 49333.7& 12 Dec 1993&   30.1& 289.0&  17.09&   1.46&    $59.4\pm2.3$& Y\\
F15-2       & 02:42:51.4& -34:19:54.3& 49333.7& 12 Dec 1993&   37.8& 282.4&  16.78&   1.59&    $60.9\pm2.3$& Y\\
F18-2       & 02:41:47.8& -34:16:37.8& 49333.7& 12 Dec 1993&   26.5& 295.7&  16.86&   1.62&    $56.7\pm2.3$& Y\\
F18-3       & 02:41:49.3& -34:12:38.0& 49333.7& 12 Dec 1993&   28.7& 302.6&  17.07&   1.48&    $66.4\pm2.6$& Y\\
F-M20       & 02:40:05.5& -34:27:43.2& 49334.6& 13 Dec 1993&    2.8& 278.9&  16.01&   2.18&    $63.7\pm2.3$& Y\\
F-M1       & 02:39:39.6& -34:19:52.2& 49334.6& 13 Dec 1993&    8.7&  17.5&  16.36&   1.95&    $56.3\pm2.3$& Y\\
       & & & 49649.8& 23 Oct 1994&    &  &  &   &    $51.8\pm2.8$& Y\\
F19-2       & 02:39:41.2& -34:11:05.7& 49334.6& 13 Dec 1993&   17.2&   7.7&  16.87&   1.49&    $56.1\pm2.2$& Y\\
F19-3       & 02:39:10.6& -34:10:54.4& 49334.6& 13 Dec 1993&   19.3&  26.6&  16.77&   1.51&    $20.1\pm2.3$& ?$^v$\\
F20-1       & 02:39:49.5& -34:05:18.6& 49334.6& 13 Dec 1993&   22.8&   1.4&  16.97&   1.52&    $57.2\pm2.7$& Y\\
F17-1       & 02:40:38.3& -33:55:54.8& 49334.7& 13 Dec 1993&   33.6& 343.5&  16.78&   1.55&   $134.0\pm2.7$& N$^v$\\
F20-2       & 02:39:08.6& -34:01:14.0& 49334.7& 13 Dec 1993&   28.4&  18.6&  16.91&   1.40&    $59.6\pm2.4$& Y\\
F20-3       & 02:40:11.7& -33:57:01.3& 49334.7& 13 Dec 1993&   31.4& 352.6&  17.02&   1.48&    $44.2\pm2.6$& Y\\
F20-4       & 02:40:28.4& -33:58:27.6& 49334.7& 13 Dec 1993&   30.6& 345.9&  16.99&   1.54&    $64.6\pm2.5$& Y\\
F17-2       & 02:40:45.6& -34:00:42.7& 49334.8& 13 Dec 1993&   29.6& 338.1&  16.76&   1.64&    $49.5\pm2.7$& Y\\
F-M2        & 02:39:43.8& -34:30:53.5& 49335.6& 14 Dec 1993&    3.2& 147.6&  16.26&   1.63&    $71.8\pm2.6$& Y\\
F13-1       & 02:41:20.8& -34:23:59.1& 49335.6& 14 Dec 1993&   18.7& 282.7&  16.80&   1.57&    $66.2\pm2.1$& Y\\
F13-2       & 02:41:39.8& -34:25:26.2& 49335.6& 14 Dec 1993&   22.3& 276.9&  19.43&   1.47&    $47.2\pm2.3$& Y\\
F13-3       & 02:42:00.3& -34:23:04.9& 49335.6& 14 Dec 1993&   26.9& 280.7&  16.78&   1.48&    $68.2\pm2.7$& Y\\
F13-4       & 02:41:56.0& -34:24:40.1& 49335.6& 14 Dec 1993&   25.7& 277.6&  16.87&   1.53&    $61.7\pm2.2$& Y\\
F11-1       & 02:41:28.6& -34:44:19.0& 49335.7& 14 Dec 1993&   25.6& 230.7&  16.74&   1.67&    $50.4\pm2.8$& Y\\
F11-2       & 02:41:20.4& -34:42:53.5& 49335.7& 14 Dec 1993&   23.4& 230.8&  16.88&   1.63&    $46.0\pm2.5$& Y\\
F13-5       & 02:41:57.7& -34:37:01.5& 49335.7& 14 Dec 1993&   27.3& 250.9&  16.85&   1.48&    $46.3\pm2.2$& Y\\
F14-1       & 02:42:22.8& -34:35:24.0& 49335.7& 14 Dec 1993&   31.8& 256.7&  16.90&   1.40&    $60.7\pm2.2$& Y\\
F11-3       & 02:40:44.0& -34:52:55.3& 49335.8& 14 Dec 1993&   27.0& 203.1&  16.82&   1.55&    $53.9\pm2.2$& Y\\
F8-1        & 02:39:32.3& -34:55:16.1& 49336.7& 15 Dec 1993&   27.4& 171.4&  16.98&   1.52&    $54.2\pm2.6$& Y\\
F10-1       & 02:41:21.4& -35:00:01.8& 49337.6& 16 Dec 1993&   36.8& 209.8&  16.94&   1.58&    $54.8\pm2.8$& Y\\
       & & & 52620.6& 12 Dec 2002& & & & &    $53.0\pm2.2$& \\
F9-2        & 02:39:35.1& -35:02:17.4& 49337.6& 16 Dec 1993&   34.3& 174.1&  17.04&   1.52&    $45.2\pm2.2$& Y\\
F9-3        & 02:39:32.2& -35:04:35.2& 49337.6& 16 Dec 1993&   36.7& 173.6&  16.94&   1.59&    $52.1\pm2.3$& Y\\
F9-4        & 02:39:17.9& -34:59:05.6& 49337.6& 16 Dec 1993&   31.7& 167.2&  16.85&   1.57&    $55.2\pm2.0$& Y\\
F9-5        & 02:38:57.8& -34:59:58.7& 49337.6& 16 Dec 1993&   33.7& 160.7&  16.77&   0.68&    $65.3\pm2.4$& N$^p$\\
F6-1        & 02:38:32.2& -35:08:30.5& 49337.7& 16 Dec 1993&   43.6& 157.9&  16.92&   1.59&    $32.9\pm2.1$& Y\\
F6-3        & 02:38:46.5& -35:01:22.5& 49337.7& 16 Dec 1993&   35.9& 157.9&  16.83&   1.50&    $86.2\pm2.8$& ?$^v$\\
F6-4        & 02:38:53.2& -34:59:02.9& 49337.7& 16 Dec 1993&   33.2& 158.6&  17.03&   1.49&    $47.8\pm2.4$& Y\\
F9-6        & 02:39:09.8& -35:17:09.9& 49337.7& 16 Dec 1993&   49.8& 170.0&  17.14&   1.37&    $68.3\pm2.5$& Y\\
        & & & 49651.9& 25 Oct 1994& & & & &    $73.0\pm2.4$& \\
F-M17       & 02:39:46.5& -34:25:52.7& 49338.5& 17 Dec 1993&    2.6&  27.8&  16.03&   1.53&    $85.1\pm2.4$& ?$^v$\\
       & & & 49651.6& 25 Oct 1994& & & & &    $86.3\pm2.4$& \\
       & & & 49652.5& 26 Oct 1994& & & & &    $85.5\pm2.5$& \\
F2-1        & 02:37:48.4& -34:36:21.9& 49338.6& 17 Dec 1993&   26.8& 108.0&  16.75&   1.58&    $50.8\pm2.6$& Y\\
        & & & 49338.7& 17 Dec 1993& & & & &    $50.1\pm2.3$& \\
F22-1       & 02:38:03.0& -34:16:39.1& 49338.6& 17 Dec 1993&   25.3&  63.1&  16.74&   1.55&    $50.5\pm2.4$& Y\\
F3-1        & 02:36:32.5& -34:28:13.8& 49338.6& 17 Dec 1993&   41.2&  90.3&  17.12&   1.47&    $54.5\pm2.7$& Y\\
F3-2        & 02:36:53.9& -34:25:52.6& 49338.6& 17 Dec 1993&   36.8&  86.7&  16.81&   1.53&    $30.5\pm2.3$& Y\\
F3-3        & 02:37:27.2& -34:30:03.3& 49338.6& 17 Dec 1993&   30.0&  93.8&  16.80&   1.63&    $45.4\pm2.5$& Y\\
F3-5        & 02:37:32.4& -34:39:53.5& 49338.6& 17 Dec 1993&   31.1& 112.4&  16.94&   1.49&    $54.8\pm2.8$& Y\\
        & & & 49338.6& 17 Dec 1993& & & & &    $55.9\pm2.5$& \\
F2-2        & 02:37:59.5& -34:34:15.7& 49338.7& 17 Dec 1993&   24.0& 104.9&  16.91&   1.54&    $67.4\pm2.6$& Y\\
F2-3        & 02:38:02.4& -34:41:00.7& 49338.7& 17 Dec 1993&   26.0& 119.8&  16.86&   1.57&    $50.9\pm2.3$& Y\\
F2-4        & 02:38:19.8& -34:38:54.1& 49338.7& 17 Dec 1993&   21.9& 119.5&  16.90&   1.51&    $47.8\pm2.7$& Y\\
F22-3       & 02:38:52.0& -34:08:39.6& 49338.7& 17 Dec 1993&   23.1&  32.6&  17.03&   1.51&    $53.1\pm2.2$& Y\\
F7-2        & 02:38:09.0& -34:51:07.1& 49338.8& 17 Dec 1993&   31.3& 137.4&  16.92&   1.58&    $51.1\pm2.8$& Y\\
F7-3        & 02:38:30.9& -34:56:54.3& 49338.8& 17 Dec 1993&   33.3& 149.9&  16.91&   1.57&    $21.4\pm2.8$& ?$^v$\\
F-M10       & 02:40:11.7& -34:31:18.8& 49339.6& 18 Dec 1993&    5.1& 231.6&  17.01&   1.58&    $56.5\pm2.2$& Y\\
F-M6        & 02:39:59.5& -34:32:43.3& 49339.6& 18 Dec 1993&    4.8& 197.9&  16.15&   1.82&    $31.9\pm2.0$& Y\\
        & & & 49652.6& 26 Oct 1994& & & & &    $37.7\pm2.7$& \\
F2-5        & 02:38:26.3& -34:25:33.7& 49339.6& 18 Dec 1993&   17.9&  81.8&  16.85&   1.54&    $51.5\pm2.3$& Y\\
F2-6        & 02:38:39.2& -34:35:54.8& 49339.6& 18 Dec 1993&   16.9& 117.4&  16.87&   1.63&    $49.0\pm2.0$& Y\\
F2-7        & 02:38:27.5& -34:30:00.6& 49339.6& 18 Dec 1993&   17.6&  96.2&  16.97&   1.51&    $60.5\pm2.5$& Y\\
F7-4        & 02:37:55.8& -34:47:16.0& 49339.7& 18 Dec 1993&   30.7& 128.7&  16.78&   1.63&    $47.4\pm2.4$& Y\\
F7-5        & 02:38:54.3& -34:43:45.6& 49339.7& 18 Dec 1993&   19.7& 142.7&  16.95&   1.44&    $54.5\pm2.6$& Y\\
F7-6        & 02:38:49.5& -34:42:29.7& 49339.7& 18 Dec 1993&   19.3& 138.0&  16.94&   1.46&    $39.2\pm2.9$& Y\\
F-M8        & 02:39:53.8& -34:29:56.5& 49340.6& 19 Dec 1993&    1.8& 190.0&  16.70&   1.59&    $51.4\pm2.3$& Y\\
        & & & 49653.6& 27 Oct 1994& & & & &    $57.0\pm2.7$& \\
F11-5       & 02:40:27.0& -34:43:42.8& 49340.6& 19 Dec 1993&   17.1& 204.6&  16.95&   1.50&    $20.2\pm2.3$& ?$^v$\\
F11-6       & 02:40:33.6& -34:45:26.3& 49340.6& 19 Dec 1993&   19.3& 206.1&  17.01&   1.52&    $75.2\pm2.5$& Y\\
F11-7       & 02:40:33.4& -34:46:57.4& 49340.6& 19 Dec 1993&   20.6& 204.1&  16.95&   1.59&    $44.6\pm2.3$& Y\\
F11-8       & 02:40:30.6& -34:52:01.3& 49340.6& 19 Dec 1993&   25.1& 198.2&  17.00&   1.50&    $61.1\pm2.7$& Y\\
F13-7       & 02:41:12.3& -34:21:53.2& 49340.7& 19 Dec 1993&   17.6& 290.7&  17.02&   1.46&    $59.2\pm2.4$& Y\\
F13-8       & 02:41:42.5& -34:21:28.5& 49340.7& 19 Dec 1993&   23.7& 286.2&  16.90&   1.46&    $72.4\pm2.6$& Y\\
F18-4       & 02:41:39.1& -34:05:50.4& 49340.7& 19 Dec 1993&   31.4& 315.2&  16.78&   1.60&    $61.2\pm2.6$& Y\\
F18-5       & 02:41:36.9& -34:10:12.4& 49340.7& 19 Dec 1993&   28.1& 309.6&  16.91&   1.51&    $43.1\pm2.8$& Y\\
F18-6       & 02:40:58.4& -34:08:02.4& 49340.8& 19 Dec 1993&   24.3& 325.8&  16.97&   1.55&    $49.7\pm2.5$& Y\\
F19-4       & 02:39:49.1& -34:09:34.2& 49341.6& 20 Dec 1993&   18.6&   2.0&  16.90&   1.48&    $36.8\pm2.7$& Y\\
F19-5       & 02:40:14.0& -34:11:07.6& 49341.6& 20 Dec 1993&   17.6& 345.2&  16.94&   1.48&    $50.8\pm2.7$& Y\\
       & & & 49341.6& 20 Dec 1993& & & & &    $42.1\pm3.0$& \\
F19-7       & 02:40:02.5& -34:03:27.2& 49341.6& 20 Dec 1993&   24.8& 355.1&  16.94&   1.51&    $54.8\pm2.3$& Y\\
F13-9       & 02:41:03.8& -34:25:27.1& 49341.7& 20 Dec 1993&   15.0& 280.3&  17.08&   1.40&    $63.9\pm2.3$& Y\\
F19-11      & 02:39:56.4& -34:12:18.6& 49341.7& 20 Dec 1993&   15.9& 356.9&  16.97&   1.45&    $73.8\pm2.8$& Y\\
F19-12      & 02:40:06.5& -34:15:52.5& 49341.7& 20 Dec 1993&   12.6& 346.6&  16.85&   1.59&    $75.5\pm2.7$& Y\\
F19-8       & 02:40:20.8& -34:16:54.2& 49341.7& 20 Dec 1993&   12.7& 332.4&  16.80&   1.47&    $79.2\pm2.3$& Y\\
F19-9       & 02:40:14.2& -34:12:23.6& 49341.7& 20 Dec 1993&   16.4& 344.0&  17.00&   1.52&    $50.9\pm2.5$& Y\\
F13-10      & 02:41:07.3& -34:25:20.5& 49341.8& 20 Dec 1993&   15.7& 280.2&  17.07&   1.45&    $65.2\pm2.6$& Y\\
F10-3       & 02:40:48.0& -35:01:23.9& 49648.6& 22 Oct 1994&   35.2& 198.9&  17.20&   1.44&    $57.6\pm2.5$& Y\\
F11-9       & 02:40:44.7& -34:41:58.2& 49648.7& 22 Oct 1994&   17.5& 217.9&  17.03&   1.51&    $58.7\pm2.5$& Y\\
F12-1       & 02:41:50.2& -34:44:41.5& 49648.7& 22 Oct 1994&   29.4& 235.6&  17.20&   1.43&    $68.9\pm2.2$& Y\\
F12-3       & 02:43:26.1& -34:55:50.0& 49648.7& 22 Oct 1994&   51.9& 237.5&  17.15&   1.53&    $ 0.6\pm2.2$& N$^v$\\
F14-3       & 02:42:27.3& -34:23:58.9& 49648.8& 22 Oct 1994&   32.2& 277.2&  17.24&   1.38&    $73.3\pm2.7$& Y\\
F15-6       & 02:43:26.1& -34:04:54.7& 49648.8& 22 Oct 1994&   49.9& 297.5&  17.29&   1.52&    $ 0.2\pm2.4$& N$^v$\\
F16-3       & 02:41:54.3& -33:52:45.7& 49648.9& 22 Oct 1994&   43.5& 324.4&  17.20&   1.43&    $-0.9\pm2.2$& N$^v$\\
F17-4       & 02:40:32.4& -34:04:04.4& 49649.6& 23 Oct 1994&   25.5& 341.0&  17.23&   1.48&    $45.1\pm2.5$& Y\\
       & & & 49652.7& 26 Oct 1994& & & & &    $55.2\pm2.8$& \\
F17-5       & 02:41:44.4& -34:01:42.4& 49649.6& 23 Oct 1994&   35.2& 318.7&  20.02&   0.54&    $48.6\pm2.7$& Y\\
F20-5       & 02:40:29.9& -33:49:37.2& 49649.7& 23 Oct 1994&   39.3& 348.5&  17.13&   1.40&    $26.7\pm2.2$& Y\\
F20-6       & 02:39:44.2& -34:05:20.0& 49649.8& 23 Oct 1994&   22.9&   4.2&  17.07&   1.48&    $28.2\pm2.2$& Y\\
F20-7       & 02:39:40.6& -33:46:44.0& 49649.8& 23 Oct 1994&   41.5&   3.4&  17.10&   1.44&    $47.6\pm2.5$& Y\\
F-M3        & 02:39:39.4& -34:28:46.7& 49650.5& 24 Oct 1994&    2.7& 103.4&  15.66&   2.04&    $57.8\pm1.8$& Y\\
F-M7        & 02:39:58.2& -34:32:05.3& 49650.6& 24 Oct 1994&    4.1& 197.2&  16.89&   1.53&    $41.5\pm3.0$& Y\\
F21-2       & 02:37:33.1& -33:59:13.0& 49650.6& 24 Oct 1994&   40.8&  45.0&  17.16&   1.52&    $50.8\pm2.1$& Y\\
F22-5       & 02:38:43.0& -34:17:19.4& 49650.7& 24 Oct 1994&   17.9&  53.0&  17.12&   1.42&    $58.0\pm2.2$& Y\\
F22-6       & 02:38:25.4& -34:10:07.8& 49650.7& 24 Oct 1994&   25.4&  45.0&  17.10&   1.48&    $66.1\pm2.6$& Y\\
F22-7       & 02:38:15.9& -34:17:18.5& 49650.8& 24 Oct 1994&   22.6&  61.5&  17.11&   1.51&    $43.8\pm2.0$& Y\\
F22-8       & 02:38:01.2& -34:18:01.4& 49650.8& 24 Oct 1994&   25.1&  66.3&  17.14&   1.46&    $68.9\pm2.6$& Y\\
F23-2       & 02:37:22.7& -34:21:31.3& 49650.8& 24 Oct 1994&   31.6&  78.1&  18.42&  -0.94&    $48.7\pm2.2$& N$^p$\\
F3-10       & 02:37:03.2& -34:25:02.5& 49650.9& 24 Oct 1994&   35.0&  85.1&  17.09&   1.47&    $71.1\pm2.8$& Y\\
F-M13       & 02:40:11.8& -34:28:54.3& 49651.5& 25 Oct 1994&    4.1& 259.4&  16.15&   1.63&    $89.2\pm2.3$& ?$^v$\\
F4-3        & 02:37:20.2& -34:46:45.9& 49651.6& 25 Oct 1994&   36.4& 120.9&  17.11&   1.48&    $58.9\pm2.7$& Y\\
F15-4       & 02:42:30.2& -34:09:36.2& 49652.6& 26 Oct 1994&   37.5& 299.4&  17.20&   1.42&    $56.2\pm2.5$& Y\\
       & & & 52623.6& 15 Dec 2002& & & & &    $54.2\pm1.7$& \\
F22-4       & 02:38:54.7& -34:20:50.6& 49652.7& 26 Oct 1994&   13.9&  58.5&  16.94&   1.58&    $45.5\pm2.6$& Y\\
F18-11      & 02:41:09.6& -34:17:21.5& 49652.8& 26 Oct 1994&   19.3& 304.0&  17.17&   1.44&    $31.7\pm2.6$& Y\\
F18-9       & 02:41:32.9& -34:16:44.5& 49652.8& 26 Oct 1994&   23.7& 298.7&  17.16&   1.46&    $61.0\pm2.9$& Y\\
F13-13      & 02:40:48.8& -34:36:43.1& 49652.9& 26 Oct 1994&   14.5& 233.6&  16.88&   1.51&    $ 1.9\pm2.7$& N$^v$\\
F24-1       & 02:35:23.1& -34:31:43.5& 49653.6& 27 Oct 1994&   55.6&  94.0&  16.83&   1.65&    $61.2\pm2.3$& Y\\
F24-2       & 02:35:02.3& -34:34:59.1& 49653.6& 27 Oct 1994&   60.1&  96.9&  16.89&   1.63&    $15.1\pm2.5$& ?$^v$\\
       & & & 52620.6& 12 Dec 2002& & & & &    $20.7\pm2.6$& \\
F26-2       & 02:36:27.2& -35:10:53.0& 49653.7& 27 Oct 1994&   60.0& 135.7&  17.00&   1.46&    $58.0\pm2.6$& Y\\
F27-2       & 02:35:01.1& -35:17:29.5& 49653.7& 27 Oct 1994&   77.5& 129.9&  17.10&   0.98&    $ 4.4\pm2.9$& N$^v$\\
F29-1       & 02:38:01.6& -35:23:10.4& 49653.7& 27 Oct 1994&   59.5& 157.7&  16.95&   1.53&    $48.4\pm2.8$& Y\\
F31-1       & 02:42:16.3& -35:02:48.7& 49653.8& 27 Oct 1994&   45.6& 220.3&  16.81&   1.62&    $ 9.2\pm2.6$& N$^v$\\
F31-3       & 02:42:07.3& -35:14:33.1& 49653.8& 27 Oct 1994&   54.0& 210.7&  17.28&   1.48&    $ 1.4\pm2.9$& N$^v$\\
       & & & 49653.8& 27 Oct 1994& & & & &     $7.5\pm2.5$& \\
       & & & 52623.6& 15 Dec 2002& & & & &     $7.5\pm2.5$& \\
F2-9        & 02:39:00.7& -34:33:17.0& 49653.9& 27 Oct 1994&   11.8& 115.9&  16.81&   1.64&    $39.7\pm2.9$& Y\\
        & & & 52623.6& 15 Dec 2002& & & & &    $55.3\pm1.8$& \\
F12-2       & 02:43:09.0& -34:45:36.3& 52620.6& 12 Dec 2002&   44.1& 246.4&  17.22&   1.48&    $78.0\pm2.0$& Y\\
F24-1139    & 02:36:09.2& -34:29:48.5& 52620.6& 12 Dec 2002&   46.0&  92.3&  17.53&   1.27&   $113.0\pm2.4$& N$^v$\\
F25-2042    & 02:35:38.8& -34:54:05.5& 52620.6& 12 Dec 2002&   58.2& 116.8&  17.60&   1.11&    $28.4\pm2.6$& Y\\
F26-4616    & 02:34:50.3& -35:16:57.2& 52620.6& 12 Dec 2002&   78.9& 128.6&  17.95&   0.96&   $123.7\pm2.6$& N$^v$\\
F29-846     & 02:38:54.3& -35:21:01.9& 52620.6& 12 Dec 2002&   54.2& 167.4&  17.86&   1.22&    $49.7\pm2.6$& Y\\
F1-32     & 02:39:33.0& -34:27:13.5& 52621.6& 13 Dec 2002&    4.1&  76.9&  16.13&   1.26&     $4.6\pm1.9$& N$^v$\\
F1-33     & 02:39:41.2& -34:32:56.4& 52621.6& 13 Dec 2002&    5.3& 154.5&  18.77&   1.18&    $52.2\pm2.4$& Y\\
F1-34     & 02:40:14.7& -34:34:16.1& 52621.6& 13 Dec 2002&    7.7& 217.0&  20.24&   0.81&    $67.8\pm2.0$& Y\\
F1-35     & 02:39:54.0& -34:42:11.0& 52621.6& 13 Dec 2002&   14.0& 181.4&  16.23&   1.86&    $34.1\pm1.7$& Y\\
F1-36     & 02:39:17.6& -34:34:38.4& 52621.6& 13 Dec 2002&    9.6& 132.3&  16.31&   1.96&    $47.5\pm2.0$& Y\\
F1-37     & 02:39:15.3& -34:18:12.8& 52621.6& 13 Dec 2002&   12.5&  37.6&  16.53&   1.78&    $51.2\pm3.2$& Y\\
F1-38     & 02:40:04.1& -34:20:10.8& 52621.6& 13 Dec 2002&    8.3& 342.9&  16.46&   1.72&    $71.9\pm1.8$& Y\\
F1-39     & 02:38:14.1& -34:17:41.2& 52621.6& 13 Dec 2002&   22.8&  62.8&  16.80&   1.54&    $41.4\pm1.8$& Y\\
F27-775     & 02:36:26.6& -35:20:03.5& 52621.6& 13 Dec 2002&   66.9& 141.1&  17.69&   1.12&    $21.5\pm3.5$& ?$^v$\\
F1-40     & 02:39:58.0& -34:35:48.6& 52622.6& 14 Dec 2002&    7.7& 188.6&  16.33&   1.80&    $48.3\pm1.7$& Y\\
F1-41     & 02:39:33.3& -34:38:30.2& 52622.6& 14 Dec 2002&   11.1& 159.3&  16.41&   1.91&    $49.8\pm1.5$& Y\\
F1-42     & 02:39:14.9& -34:35:37.3& 52622.6& 14 Dec 2002&   10.7& 134.2&  16.27&   1.84&    $38.0\pm1.9$& Y\\
F1-43     & 02:39:16.6& -34:32:35.2& 52622.6& 14 Dec 2002&    8.6& 121.1&  20.09&   0.84&    $59.5\pm1.8$& Y\\
F1-44     & 02:38:43.5& -34:22:56.7& 52622.6& 14 Dec 2002&   15.1&  69.9&  16.94&   1.36&    $81.0\pm2.9$& Y\\
F1-45     & 02:39:21.2& -34:23:32.8& 52622.6& 14 Dec 2002&    7.9&  54.4&  19.94&   1.02&    $65.4\pm1.8$& Y\\
F1-46     & 02:39:31.4& -34:22:58.7& 52622.6& 14 Dec 2002&    6.7&  39.8&  16.28&   1.88&    $56.1\pm1.5$& Y\\
F1-47     & 02:39:49.8& -34:27:30.3& 52622.6& 14 Dec 2002&    0.8&  39.0&  16.33&   1.88&    $48.5\pm1.7$& Y\\
F1-48     & 02:40:16.0& -34:23:19.1& 52622.6& 14 Dec 2002&    6.9& 314.6&  16.60&   1.64&    $55.8\pm1.7$& Y\\
F14-1805    & 02:42:08.9& -34:33:07.3& 52622.6& 14 Dec 2002&   28.6& 259.8&  17.03&   1.35&    $51.5\pm2.0$& Y\\
F15-2830    & 02:43:18.8& -34:07:24.0& 52622.6& 14 Dec 2002&   47.4& 295.7&  17.31&   1.34&    $51.7\pm2.8$& Y\\
F16-4010    & 02:42:52.0& -33:54:55.2& 52622.6& 14 Dec 2002&   49.9& 311.6&  17.08&   1.28&    $55.1\pm2.4$& Y\\
F1-49     & 02:39:32.7& -34:31:00.7& 52623.6& 15 Dec 2002&    5.0& 125.3&  17.23&   1.49&    $60.1\pm2.0$& Y\\
F1-50     & 02:39:24.4& -34:32:53.1& 52623.6& 15 Dec 2002&    7.5& 129.5&  16.16&   1.71&    $40.3\pm1.6$& Y\\
F12-451     & 02:41:59.5& -34:43:34.2& 52623.6& 15 Dec 2002&   30.4& 239.3&  17.41&   1.32&    $31.0\pm2.0$& Y\\
F21-3329    & 02:38:16.1& -34:02:05.9& 52623.6& 15 Dec 2002&   32.8&  37.5&  17.04&   1.37&    $58.2\pm2.1$& Y\\
F31-1198    & 02:42:30.4& -35:09:37.6& 52623.6& 15 Dec 2002&   52.7& 217.8&  16.94&   1.19&    $16.8\pm1.9$& ?$^v$\\
F31-365     & 02:42:31.0& -35:00:37.1& 52623.6& 15 Dec 2002&   46.0& 224.9&  17.17&   1.17&    $67.5\pm2.0$& Y\\
F9-7731     & 02:39:08.9& -35:07:10.7& 52623.6& 15 Dec 2002&   40.0& 167.2&  17.09&   1.51&    $48.5\pm1.7$& Y\\
F9-8025     & 02:39:02.3& -35:09:13.6& 52623.6& 15 Dec 2002&   42.3& 166.0&  17.22&   1.41&    $69.2\pm1.9$& Y\\
CPD-35$^{\circ}$919         & 02:39:35.2& -34:30:37.0& 48955.6& 29 Nov 1992&    4.3& 125.1& & &     $4.6\pm0.7$& N$^{v,p}$\\
         & & & 48956.5& 30 Nov 1992& & & & &     $5.1\pm0.7$& \\
         & & & 48957.6& 01 Dec 1992& & & & &     $2.8\pm0.8$& \\
         & & & 48958.6& 02 Dec 1992& & & & &     $3.5\pm0.8$& \\
         & & & 48959.6& 03 Dec 1992& & & & &     $3.4\pm0.7$& \\
         & & & 48961.6& 05 Dec 1992& & & & &     $4.6\pm0.8$& \\
         & & & 48962.6& 06 Dec 1992& & & & &     $5.0\pm0.8$& \\
         & & & 48963.6& 07 Dec 1992& & & & &     $4.5\pm0.8$& \\
         & & & 49333.6& 12 Dec 1993& & & & &     $3.8\pm0.7$& \\
         & & & 49333.8& 12 Dec 1993& & & & &     $3.7\pm0.8$& \\
         & & & 49334.5& 13 Dec 1993& & & & &     $3.1\pm0.8$& \\
         & & & 49334.8& 13 Dec 1993& & & & &     $3.8\pm0.8$& \\
         & & & 49335.5& 14 Dec 1993& & & & &     $3.4\pm0.8$& \\
         & & & 49335.8& 14 Dec 1993& & & & &     $6.4\pm0.9$& \\
         & & & 49336.7& 15 Dec 1993& & & & &     $3.2\pm0.7$& \\
         & & & 49337.5& 16 Dec 1993& & & & &     $3.9\pm0.7$& \\
         & & & 49337.8& 16 Dec 1993& & & & &     $3.7\pm0.7$& \\
         & & & 49338.5& 17 Dec 1993& & & & &     $1.9\pm0.7$& \\
         & & & 49339.5& 18 Dec 1993& & & & &     $6.0\pm0.7$& \\
         & & & 49339.8& 18 Dec 1993& & & & &     $4.8\pm0.7$& \\
         & & & 49340.5& 19 Dec 1993& & & & &     $5.3\pm0.7$& \\
         & & & 49340.8& 19 Dec 1993& & & & &     $4.1\pm0.7$& \\
         & & & 49341.5& 20 Dec 1993& & & & &     $3.2\pm0.6$& \\
         & & & 49341.8& 20 Dec 1993& & & & &     $4.0\pm0.7$& \\
         & & & 49648.5& 22 Oct 1994& & & & &     $3.7\pm0.7$& \\
         & & & 49649.5& 23 Oct 1994& & & & &     $4.2\pm0.7$& \\
         & & & 49650.5& 24 Oct 1994& & & & &     $3.9\pm0.7$& \\
         & & & 49651.5& 25 Oct 1994& & & & &     $4.4\pm0.8$& \\
         & & & 49652.5& 26 Oct 1994& & & & &     $4.0\pm0.8$& \\
         & & & 49652.9& 26 Oct 1994& & & & &     $3.6\pm0.8$& \\
         & & & 49653.5& 27 Oct 1994& & & & &     $4.9\pm0.8$& \\
         & & & 49653.8& 27 Oct 1994& & & & &     $3.7\pm0.9$& \\
  \enddata
\end{deluxetable*}

\begin{deluxetable*}{lccccccc}
  \tabletypesize{\scriptsize}
  \tablenum{2}
  \tablewidth{0pc}
  \tablecaption{\scriptsize Results for radial velocity standard stars
    \label{tab:standards}}
  \tablehead{\colhead{Star} & \colhead{$\alpha_{2000}$} & \colhead{$\delta_{2000}$} & \colhead{v$_{published}$\tablenotemark{a}} & \colhead{HJD} & \colhead{UT Date} & \colhead{$v$} & \colhead{$\langle v \rangle$\tablenotemark{b}}\\
    \colhead{}&\colhead{}&\colhead{}&\colhead{(km s$^{-1}$)}&\colhead{(-2400000.0)}&\colhead{of observation}&\colhead{(km s$^{-1}$)}&\colhead{(km s$^{-1}$)}
  }
  \startdata
  HD196983    & 20:41:50.5& -33:53:16.9& $-9.1\pm0.3$\tablenotemark{1}& 48954.5& 28 Nov 1992&    $-9.0\pm0.6$ & $-9.3\pm0.9$\\
  &  & & & 48955.5& 29 Nov 1992&     $-9.3\pm0.7$\\
  &  & & & 48956.5& 30 Nov 1992&     $-8.1\pm0.6$\\
  &  & & & 48957.5& 01 Dec 1992&    $-10.1\pm0.8$\\
  &  & & & 48958.5& 02 Dec 1992&    $-10.3\pm0.8$\\
  &  & & & 48959.5& 03 Dec 1992&     $-9.9\pm0.8$\\
  &  & & & 48960.5& 04 Dec 1992&     $-9.2\pm0.8$\\
  &  & & & 48961.5& 05 Dec 1992&     $-9.9\pm0.8$\\
  &  & & & 48962.5& 06 Dec 1992&     $-9.8\pm0.8$\\
  &  & & & 48963.5& 07 Dec 1992&     $-9.8\pm0.7$\\
  &  & & & 49648.5& 22 Oct 1994&     $-9.9\pm0.7$\\
  &  & & & 49649.5& 23 Oct 1994&     $-8.7\pm0.7$\\
  &  & & & 49650.5& 24 Oct 1994&     $-8.6\pm0.7$\\
  &  & & & 49651.5& 25 Oct 1994&     $-8.3\pm0.8$\\
  &  & & & 49652.5& 26 Oct 1994&     $-9.4\pm0.8$\\
  &  & & & 49653.5& 27 Oct 1994&     $-9.1\pm0.7$\\
  \\
  HD219509    & 23:17:17.6& -66:54:48.4& $+67.5\pm0.5$\tablenotemark{1}& 48954.5& 28 Nov 1992&    $67.9\pm0.8$ & $+67.8\pm1.0$\\
  &  & & & 48955.5& 29 Nov 1992&     $67.0\pm0.8$\\
  &  & & & 48956.5& 30 Nov 1992&     $70.0\pm0.8$\\
  &  & & & 48957.5& 01 Dec 1992&     $68.5\pm0.9$\\
  &  & & & 48958.5& 02 Dec 1992&     $67.0\pm0.9$\\
  &  & & & 48959.5& 03 Dec 1992&     $68.6\pm0.9$\\
  &  & & & 48960.5& 04 Dec 1992&     $68.3\pm0.9$\\
  &  & & & 48961.5& 05 Dec 1992&     $67.1\pm0.9$\\
  &  & & & 48962.5& 06 Dec 1992&     $67.8\pm0.9$\\
  &  & & & 48963.5& 07 Dec 1992&     $68.5\pm0.9$\\
  &  & & & 49333.5& 12 Dec 1993&     $66.6\pm0.9$\\
  &  & & & 49335.5& 14 Dec 1993&     $68.1\pm1.0$\\
  &  & & & 49337.5& 16 Dec 1993&     $67.2\pm0.8$\\
  &  & & & 49338.5& 17 Dec 1993&     $68.0\pm0.8$\\
  &  & & & 49339.5& 18 Dec 1993&     $68.2\pm0.9$\\
  &  & & & 49340.5& 19 Dec 1993&     $66.7\pm0.8$\\
  &  & & & 49341.5& 20 Dec 1993&     $67.9\pm0.8$\\
  &  & & & 49648.5& 22 Oct 1994&     $66.1\pm0.9$\\
  &  & & & 49649.5& 23 Oct 1994&     $67.9\pm0.8$\\
  &  & & & 49650.5& 24 Oct 1994&     $68.2\pm0.8$\\
  &  & & & 49651.5& 25 Oct 1994&     $67.6\pm0.9$\\
  &  & & & 49652.5& 26 Oct 1994&     $68.0\pm0.9$\\
  \\
  CPD-432527  & 06:32:15.3& -43:31:14.3& $+19.7\pm0.9$\tablenotemark{1}& 48954.9& 28 Nov 1992&    $19.7\pm0.7$ & $+19.8\pm0.9$\\
  & & & & 48955.9& 29 Nov 1992&     $20.4\pm0.7$\\
  & & & & 48957.8& 01 Dec 1992&     $19.8\pm0.8$\\
  & & & & 48958.9& 02 Dec 1992&     $20.1\pm0.8$\\
  & & & & 48959.9& 03 Dec 1992&     $20.2\pm0.8$\\
  & & & & 48960.9& 04 Dec 1992&     $18.1\pm0.8$\\
  & & & & 48961.9& 05 Dec 1992&     $20.0\pm0.9$\\
  & & & & 48962.9& 06 Dec 1992&     $20.0\pm0.8$\\
  & & & & 48963.9& 07 Dec 1992&     $20.0\pm0.8$\\
  & & & & 49333.8& 12 Dec 1993&     $19.7\pm0.8$\\
  & & & & 49334.9& 13 Dec 1993&     $20.0\pm0.8$\\
  & & & & 49335.9& 14 Dec 1993&     $22.1\pm0.9$\\
  & & & & 49335.9& 14 Dec 1993&     $20.0\pm0.7$\\
  & & & & 49336.9& 15 Dec 1993&     $18.7\pm0.7$\\
  & & & & 49338.9& 17 Dec 1993&     $20.0\pm0.7$\\
  & & & & 49339.9& 18 Dec 1993&     $19.4\pm0.7$\\
  & & & & 49340.9& 19 Dec 1993&     $20.4\pm0.7$\\
  & & & & 49341.9& 20 Dec 1993&     $20.4\pm0.8$\\
  & & & & 49648.9& 22 Oct 1994&     $19.7\pm0.7$\\
  & & & & 49649.9& 23 Oct 1994&     $19.7\pm0.7$\\
  & & & & 49650.9& 24 Oct 1994&     $20.6\pm0.7$\\
  & & & & 49651.9& 25 Oct 1994&     $19.1\pm0.8$\\
  & & & & 49652.9& 26 Oct 1994&     $18.7\pm0.9$\\
  & & & & 49653.9& 27 Oct 1994&     $18.9\pm0.8$\\
  \\
  HD23214     & 03:42:09.1& -34:25:14.8& $-4.3\pm1.8$\tablenotemark{2}& 49333.6& 12 Dec 1993&    $-4.9\pm0.7$& $-5.1\pm0.9$\\
  &   & & & 49334.5& 13 Dec 1993&     $-5.3\pm0.7$\\
  &   & & & 49336.8& 15 Dec 1993&     $-5.3\pm0.7$\\
  &   & & & 49648.9& 22 Oct 1994&     $-5.0\pm0.8$\\
  \\
  HD43880     & 06:17:06.3& -34:44:13.1& $+43.6\pm2.4$\tablenotemark{2}& 49333.8& 12 Dec 1993&    $45.3\pm0.8$ & $+46.3\pm0.9$\\
  &   & & & 49334.8& 13 Dec 1993&     $47.9\pm0.8$\\
  &   & & & 49335.8& 14 Dec 1993&     $47.4\pm0.8$\\
  &   & & & 49336.8& 15 Dec 1993&     $45.1\pm0.7$\\
  &   & & & 49337.9& 16 Dec 1993&     $45.9\pm0.7$\\
  &   & & & 49338.8& 17 Dec 1993&     $47.4\pm0.6$\\
  &   & & & 49339.9& 18 Dec 1993&     $46.0\pm0.7$\\
  &   & & & 49340.9& 19 Dec 1993&     $45.9\pm0.7$\\
  &   & & & 49341.9& 20 Dec 1993&     $45.8\pm0.7$\\
  \\
  Twilight sky& \nodata& \nodata& \nodata& 49333.9&  12 Dec 1993&    $-1.4\pm1.4$ & $-0.2\pm1.2$\\
 & & & & 49334.5& 13 Dec 1993&     $-0.6\pm1.0$\\
 & & & & 49335.5& 14 Dec 1993&      $0.7\pm1.1$\\
 & & & & 49335.5& 14 Dec 1993&      $0.2\pm1.0$\\
 & & & & 49337.5& 16 Dec 1993&     $-0.8\pm1.0$\\
 & & & & 49338.5& 17 Dec 1993&     $-0.3\pm1.0$\\
 & & & & 49339.5& 18 Dec 1993&     $-0.3\pm1.0$\\
 & & & & 49341.5& 20 Dec 1993&      $0.0\pm1.0$\\
  \\
  HD6655      & 01:05:18.0& -72:33:21.0& $+19.5\pm0.3$\tablenotemark{1}& 52620.5& 12 Dec 2002&    $18.8\pm0.7$& $+19.2\pm1.1$\\
  &    & & & 52620.5& 12 Dec 2002&     $18.9\pm0.7$\\
  &    & & & 52621.5& 13 Dec 2002&     $20.0\pm0.7$\\
  &    & & & 52622.5& 14 Dec 2002&     $19.4\pm0.7$\\
  \\
  HD21581     & 03:28:54.8&  -00:25:03.1& $+154\pm1$\tablenotemark{3}& 52620.6& 12 Dec 2002&   $152.1\pm0.8$ & $+151.3\pm1.3$\\
  &   & & & 52621.6& 13 Dec 2002&    $150.6\pm0.9$\\
  \\
  SAO217998   & 06:32:15.6& -43:31:13.4& $+13.1$\tablenotemark{4}& 52620.6& 12 Dec 2002&    $18.9\pm0.5$ & $+19.0\pm1.6$\\
  & & & & 52621.6& 13 Dec 2002&     $18.4\pm0.7$\\
  & & & & 52621.8& 13 Dec 2002&     $19.6\pm0.5$\\
  & & & & 52622.6& 14 Dec 2002&     $19.1\pm0.5$\\
  & & & & 52622.8& 14 Dec 2002&     $17.8\pm0.6$\\
  & & & & 52623.5& 15 Dec 2002&     $20.1\pm0.6$\\
  & & & & 52623.6& 15 Dec 2002&     $18.0\pm0.6$\\
  & & & & 52623.6& 15 Dec 2002&     $20.2\pm0.6$\\
  & & & & 52623.8& 15 Dec 2002&     $19.4\pm0.6$\\
  \\
  HD83516     & 09:38:02.9& -35:04:34.0& $+43.5\pm0.2$\tablenotemark{1}& 52620.9& 12 Dec 2002&    $43.2\pm0.4$ & $+42.6\pm0.7$\\
  &   & & & 52621.8& 13 Dec 2002&     $42.4\pm0.4$\\
  &   & & & 52622.9& 14 Dec 2002&     $42.9\pm0.5$\\
  &   & & & 52622.9& 14 Dec 2002&     $42.6\pm0.4$\\
  &   & & & 52622.9& 14 Dec 2002&     $43.9\pm0.6$\\
  &   & & & 52623.8& 15 Dec 2002&     $40.2\pm0.6$\\
  \\
  HD48381     & 06:41:43.0& -33:28:13.2& $+40.5\pm0.2$\tablenotemark{1}& 52621.8& 13 Dec 2002&    $40.2\pm0.5$ & $+40.2\pm0.9$\\
  &   & & & 52623.8& 15 Dec 2002&     $40.3\pm0.6$\\
  \\
  SAO201636   & 10:41:09.5& -30:47:05.8& $+262\pm1$\tablenotemark{3}& 52621.8& 13 Dec 2002&   $264.3\pm0.9$ & $+264.3\pm1.4$\\
  \\
  HD2796      & 00:31:16.5& -16:47:43.6& $-61\pm1$\tablenotemark{3}& 52623.5& 15 Dec 2002&   $-64.5\pm1.0$ & $-54.5\pm1.5$\\
  \\
  HD103545    & 11:55:26.2&  09:07:54.4& $+180\pm1$\tablenotemark{3}& 52623.9& 15 Dec 2002&   $177.0\pm1.0$ & $+177.0\pm1.5$\\
  \\
  \enddata
  \tablenotetext{a}{published radial velocity}
  \tablenotetext{b}{weighted mean measured radial velocity from the results presented in this table}
  \tablenotetext{1}{reference: CORAVEL study (Udry et al. 1999)}
  \tablenotetext{2}{reference: Olszewski et al. (1991)}
  \tablenotetext{3}{reference: Beers et al. (2000)}
  \tablenotetext{4}{reference: Evans (1967)}
  
\end{deluxetable*}

\begin{deluxetable*}{lcccc}
  \tabletypesize{\scriptsize}
  \tablenum{3}
  \tablewidth{0pc}
  \tablecaption{\scriptsize Repeat velocity measurements.   
  \label{tab:repeats}}
  \tablehead{\colhead{Star}& \colhead{$\langle v \rangle$}& \colhead{$\chi^2$\tablenotemark{a}}& \colhead{p($\chi^2$)\tablenotemark{b}}& \colhead{N} \\
    \colhead{}&\colhead{(km s$^{-1}$)}&\colhead{}& \colhead{}& \colhead{}
  }
  \startdata
  CPD-432527&       $19.8\pm0.8$&    21.8&    0.5327&  24\\
  Twilight Sky&     $-0.2\pm1.0$&     1.3&    0.9722&   7\\
  F1-4&          $43.7\pm1.8$&      0.2&    0.9204&   3\\
  F1-2&          $44.2\pm2.4$&      0.4&    0.5105&   2\\
  F1-1&          $61.4\pm2.3$&      6.6&    0.0103&   2\\
  CPD-35$^{\circ}$919&  $4.1\pm0.7$&     47.6&    0.0288&  32\\
  F-M17&            $85.6\pm2.4$&      0.1&    0.9364&   3\\
  F-M26&            $42.0\pm2.2$&      3.6&    0.3080&   3\\
  F-M4&             $52.7\pm2.2$&      0.1&    0.7369&   2\\
  F-M6&             $34.0\pm2.2$&      3.1&    0.0785&   2\\
  F-M8&             $53.8\pm2.5$&      2.6&    0.1101&   2\\
  F10-1&            $53.7\pm2.4$&      0.3&    0.6086&   2\\
  F15-4&            $54.8\pm1.9$&      0.4&    0.5157&   2\\
  F17-4&            $49.5\pm2.6$&      7.3&    0.0069&   2\\
  F19-5&            $46.9\pm2.8$&      4.7&    0.0295&   2\\
  F2-1&             $50.4\pm2.4$&      0.0&    0.8516&   2\\
  F2-9&             $50.9\pm2.1$&    21.4&    0.0000&   2\\
  F24-2&            $17.7\pm2.5$&     2.5&    0.1137&   2\\
  F3-5&             $55.4\pm2.6$&      0.1&    0.7647&   2\\
  F31-3&            $ 5.9\pm2.6$&      3.2&    0.2063&   3\\
  F9-6&             $70.7\pm2.5$&      1.9&    0.1734&   2\\
  HD196983&         $-9.3\pm0.7$&     12.7&    0.6242&  16\\
  HD21581&         $151.3\pm0.8$&      1.5&    0.2151&   2\\
  HD219509&         $67.8\pm0.9$&     22.2&    0.3862&  22\\
  HD23214&          $-5.1\pm0.7$&      0.3&    0.9608&   4\\
  HD43880&          $46.3\pm0.7$&     14.6&    0.0685&   9\\
  HD48381&          $40.2\pm0.5$&      0.0&    0.8648&   2\\
  HD6655&           $19.2\pm0.7$&      1.9&    0.5916&   4\\
  HD83516&          $42.6\pm0.5$&     23.6&    0.0003&   6\\
  SAO217998&        $19.0\pm1.0$&     17.0&    0.0306&   9\\
  \enddata
  \tablenotetext{a}{$\chi^2=\displaystyle\sum_{i=1}^N\frac{(v_{i}-\langle v \rangle)^2}{\sigma_i^2}$}
  \tablenotetext{b}{Probability that $\chi^2$ would have at least the measured value, given N independent measurements having the estimated uncertainties.}
\end{deluxetable*}

\begin{deluxetable*}{lccccc}
  \tabletypesize{\scriptsize}
  \tablenum{4}
  \tablewidth{0pc}
  \tablecaption{\scriptsize Comparison with previously measured velocities.  Entries with UT date prior to 1992 were published in Mateo et al. (1991).  All other entries are from the present study. 
    \label{tab:compare}}
  \tablehead{\colhead{Star}& \colhead{UT Date}& \colhead{$v$}& \colhead{$\langle v \rangle$} & \colhead{$\chi^2$} & \colhead{p$(\chi^2)$} \\
\colhead{}& \colhead{of observation}& \colhead{km s$^{-1}$}& \colhead{km s$^{-1}$} & \colhead{} & \colhead{} \\
  }
  \startdata
F-M1  & 30 Nov 1989&   $56.7\pm2.9$&  $55.1\pm2.6$&     2.0 &    0.3734 \\
  & 13 Dec 1993&   $56.3\pm2.3$&   & &  \\
  & 23 Oct 1994&   $51.8\pm2.8$&  & &  \\
\\
F-M10 & 03 Dec 1989&   $65.1\pm1.3$&  $63.1\pm1.6$&     5.8 &    0.0562 \\
 & 17 Nov 1990&   $63.8\pm1.8$&  & &  \\
 & 18 Dec 1993&   $56.5\pm2.2$&  & &  \\
\\
F-M13 & 03 Dec 1989&   $92.1\pm2.0$&  $90.9\pm2.1$&     0.9 &    0.3414 \\
 & 25 Oct 1994&   $89.2\pm2.3$&  & & \\
\\
F-M15 & 04 Dec 1989&   $57.9\pm2.0$&  $60.0\pm1.9$&     2.0 &    0.1579 \\
 & 12 Dec 1993&   $61.7\pm1.8$&   & & \\
\\
F-M17 & 05 Dec 1989&   $87.8\pm1.9$&  $86.5\pm2.2$&     0.2 &    0.9931 \\
 & 14 Nov 1990&   $86.7\pm2.2$&   & & \\
 & 17 Dec 1993&   $85.1\pm2.4$&   & & \\
 & 25 Oct 1994&   $86.3\pm2.4$&   & & \\
 & 26 Oct 1994&   $85.5\pm2.5$&   & & \\
\\
F-M18 & 05 Dec 1989&   $54.3\pm2.4$&  $54.8\pm2.6$&     0.1 &    0.8231 \\
 &  05 Dec 1992&   $55.0\pm2.8$&   & & \\
\\
F-M2  & 01 Dec 1989&   $71.6\pm1.5$&  $71.6\pm1.8$&     0.0 &    0.9469 \\
  & 14 Dec 1993&   $71.8\pm2.6$&   & & \\
\\
F-M20 & 15 Nov 1990&   $38.9\pm2.0$&  $49.6\pm2.1$&   66.2 &    0.0000 \\
 & 13 Dec 1993&   $63.7\pm2.3$& &   & \\
\\
F-M26 & 18 Nov 1990&   $40.3\pm1.6$&  $41.4\pm1.9$&      2.1 &    0.54996 \\
 & 07 Dec 1992&   $41.2\pm2.3$& &   & \\
 & 20 Dec 1993&   $43.8\pm2.0$& &   & \\
 & 23 Oct 1994&   $40.3\pm2.5$& &   & \\
\\
F-M3  & 01 Dec 1990&   $59.1\pm2.8$&  $58.2\pm2.1$&     0.2 &    0.6961 \\
  & 24 Oct 1994&   $57.8\pm1.8$&  &  & \\
\\
F-M4  & 01 Dec 1989&   $60.0\pm3.6$&  $53.6\pm2.3$&     1.2 &    0.7451 \\
  & 15 Nov 1990&   $52.9\pm2.1$&  &  & \\
  & 12 Dec 1993&   $52.3\pm2.0$&  &  & \\
  & 22 Oct 1994&   $53.4\pm2.6$&  &  & \\
\\
F-M6  & 02 Dec 1989&   $34.9\pm1.9$&  $34.5\pm2.2$&     1.1 &    0.7805 \\
  & 16 Nov 1990&   $35.5\pm2.9$&  &  & \\
  & 18 Dec 1993&   $31.9\pm2.0$&  &  & \\
  & 26 Oct 1994&   $37.7\pm2.7$&  & & \\
\\
F-M7  & 02 Dec 1989&   $42.3\pm2.5$&  $42.0\pm2.7$&     0.0 &    0.8377 \\
  & 24 Oct 1994&   $41.5\pm3.0$& & &  \\
\\
F-M8  & 02 Dec 1989&   $54.4\pm2.5$&  $54.6\pm2.5$&     1.2 &    0.7644 \\
  & 04 Dec 1989&   $56.9\pm2.7$& & &  \\
  & 19 Dec 1993&   $51.4\pm2.3$& & &  \\
  & 27 Oct 1994&   $57.0\pm2.7$& & &  \\
  \enddata
\end{deluxetable*}

\begin{deluxetable*}{lccccc}
  \tabletypesize{\scriptsize}
  \tablenum{5}
  \tablewidth{0pc}
   \label{tab:rotation}
  \tablecaption{\scriptsize Rotation signals in the galactic rest frame.}
  \tablehead{\colhead{Source}& \multicolumn{2}{c}{Proper Motion}&\multicolumn{2}{c}{GRF Rotation}&\colhead{Significance}\\
    \colhead{}&\colhead{$\mu_l$}&\colhead{$\mu_b$}&\colhead{Speed}&\colhead{Axis}&\colhead{}\\
    \colhead{}&\colhead{(mas century$^{-1}$)}&\colhead{(mas century$^{-1}$)}&  \colhead{(km s$^{-1}$)}& \colhead{($^{\circ}$)} & \colhead{}\\
  }
  \startdata
  Piatek et al. (2002)& $32\pm13$& $33\pm13$& $2.5\pm0.4$ & $112\pm8$&10\%\\
  Dinescu et al. (2004)& $-13\pm16$ & $34\pm16$& $1.2\pm0.4$& $116\pm12$&74\%\\
  & -52 & +41 & 0.2 & \nodata& 100\%\\
  \enddata
\end{deluxetable*}

\begin{deluxetable*}{ccccccccccc}
  \tabletypesize{\scriptsize}
  \tablenum{6}
  \tablewidth{0pc}
  \label{tab:king2}
  \tablecaption{\scriptsize Parameters and derived values from example two-component King models.}
  \tablehead{\colhead{N\tablenotemark{a}}&\colhead{$[M/L]_L$\tablenotemark{b}}&\colhead{$W_0$}&\colhead{$\rho_{0D}/\rho_{0L}$}&\colhead{$r_{cD}/r_{cL}$}&\colhead{$r_s$}&\colhead{$v_s$}&\colhead{$\chi^2$/dof}&\colhead{$\rho_{0D}$}&\colhead{$M_D$\tablenotemark{c}}&\colhead{$M/L$\tablenotemark{d}}\\
    \colhead{}&\colhead{$([M/L]_{\sun})$}&\colhead{}&\colhead{}&\colhead{}&\colhead{(pc)}&\colhead{(km s$^{-1}$)}&\colhead{}&\colhead{($M_{\sun}$ pc$^{-3}$)}&\colhead{($10^8 M_{\sun}$)}&\colhead{($[M/L]_{\sun}$)}}
  \startdata
176 & 1 & 3.3 & 0.0 & \nodata & 512 & 5.5 & 19 & 0.0 & 0.00; 0.00 & 1; 1\\
176 & 1 & 3.3 & 3.5 & 1.7 & 772 & 17.5 & 2.2 & 0.067 & 2.3; 2.2 & 17; 16\\
176 & 1 & 5.0 & 3.2 & 2.1 & 775 & 17.0 & 1.6 & 0.061 & 4.4; 3.7 & 31; 27\\
176 & 1 & 7.0 & 3.1 & 2.2 & 732 & 15.8 & 1.5 & 0.059 & 6.5; 4.1 & 46; 30\\
176 & 1 & 9.0 & 3.2 & 2.3 & 722 & 15.7 & 1.6 & 0.060 & 11.0; 4.4 & 74; 31\\
\\
176 & 2 & 3.3 & 0.0 & \nodata & 512 & 7.8 & 13 & 0.31 & 0.00; 0.00 & 2; 2\\
176 & 2 & 3.3 & 1.4 & 1.8 & 729 & 17.1 & 2.5 & 0.053 & 2.0; 2.0 & 15; 15\\ 
176 & 2 & 5.0 & 1.2 & 2.3 & 717 & 16.1 & 1.7 & 0.046 & 3.8; 3.4 & 28; 26\\ 
176 & 2 & 7.0 & 1.2 & 2.6 & 683 & 15.3 & 1.6 & 0.046 & 5.8; 4.4 & 42; 32\\ 
176 & 2 & 9.0 & 1.2 & 2.5 & 639 & 14.4 & 1.5 & 0.046 & 7.4; 4.3 & 52; 32\\ 
\\
176 & 3 & 3.3 & 0.0 & \nodata & 512 & 9.6 & 10 & 0.47 & 0.00; 0.00 & 3; 3\\ 
176 & 3 & 3.3 & 0.8 & 1.5 & 632 & 15.8 & 3.4 & 0.046 & 1.2; 1.2 & 12; 12\\ 
176 & 3 & 5.0 & 0.7 & 2.0 & 610 & 14.8 & 2.0 & 0.040 & 2.4; 2.3 & 18; 18\\ 
176 & 3 & 7.0 & 0.6 & 2.5 & 589 & 13.9 & 1.7 & 0.035 & 4.1; 3.2 & 27; 24\\ 
176 & 3 & 9.0 & 0.6 & 3.1 & 601 & 14.1 & 1.7 & 0.035 & 7.1; 4.8 & 51; 36\\ 
\\
182 & 1 & 3.3 & 4.3 & 1.8 & 852 & 20.9 & 2.3 & 0.082 & 3.8; 3.6 & 27; 26\\ 
182 & 1 & 5.0 & 4.2 & 1.9 & 760 & 18.5 & 1.8 & 0.080 & 5.1; 4.2 & 35; 30\\ 
182 & 1 & 7.0 & 4.1 & 2.2 & 766 & 18.5 & 1.7 & 0.078 & 9.3; 5.6 & 67; 41\\ 
182 & 1 & 9.0 & 4.3 & 2.1 & 709 & 17.4 & 1.7 & 0.082 & 14.6; 5.9 & 98; 35\\
\\ 
182 & 2 & 3.3 & 1.8 & 1.9 & 791 & 20.0 & 2.5 & 0.069 & 3.2; 3.1 & 24; 24\\ 
182 & 2 & 5.0 & 1.7 & 2.1 & 726 & 18.1 & 2.0 & 0.065 & 4.7; 4.2 & 34; 30\\ 
182 & 2 & 7.0 & 1.6 & 2.4 & 699 & 17.1 & 1.7 & 0.061 & 7.2; 5.2 & 52; 38\\ 
182 & 2 & 9.0 & 1.6 & 2.6 & 681 & 16.6 & 1.6 & 0.061 & 10.5; 5.8 & 76; 42\\ 
\\
182 & 3 & 3.3 & 1.1 & 1.7 & 681 & 18.3 & 3.3 & 0.064 & 2.0; 2.0 & 15; 15\\ 
182 & 3 & 5.0 & 0.9 & 2.1 & 654 & 16.8 & 2.1 & 0.052 & 3.6; 3.3 & 27; 24\\ 
182 & 3 & 7.0 & 0.8 & 2.7 & 644 & 16.0 & 1.7 & 0.046 & 6.1; 4.8 & 45; 36\\ 
182 & 3 & 9.0 & 0.8 & 2.9 & 620 & 15.5 & 1.6 & 0.046 & 8.5; 5.4 & 60; 39\\ 
\\
186 & 1 & 3.3 & 5.1 & 1.8 & 857 & 22.6 & 2.1 & 0.097 & 4.4; 4.2 & 32; 30\\ 
186 & 1 & 5.0 & 4.9 & 2.0 & 778 & 20.2 & 1.7 & 0.093 & 6.2; 5.0 & 43; 36\\ 
186 & 1 & 7.0 & 4.8 & 2.1 & 750 & 19.3 & 1.5 & 0.092 & 10.0; 5.8 & 70; 42\\ 
186 & 1 & 9.0 & 5.0 & 2.1 & 722 & 18.9 & 1.6 & 0.095 & 18.0; 5.9 & 120; 41\\ 
\\
186 & 2 & 3.3 & 2.1 & 2.0 & 835 & 22.2 & 2.2 & 0.080 & 4.2; 4.1 & 32; 30\\ 
186 & 2 & 5.0 & 2.1 & 2.1 & 736 & 19.6 & 1.8 & 0.080 & 5.6; 4.8 & 40; 36\\ 
186 & 2 & 7.0 & 2.0 & 2.2 & 694 & 18.2 & 1.6 & 0.076 & 8.0; 5.6 & 56; 40\\ 
186 & 2 & 9.0 & 2.0 & 2.4 & 684 & 17.9 & 1.5 & 0.076 & 12.6; 6.2 & 86; 44\\
\\
186 & 3 & 3.3 & 1.3 & 1.8 & 725 & 20.4 & 2.7 & 0.075 & 2.8; 2.8 & 21; 21\\ 
186 & 3 & 5.0 & 1.1 & 2.3 & 703 & 18.9 & 1.9 & 0.063 & 5.1; 4.6 & 36; 36\\ 
186 & 3 & 7.0 & 1.1 & 2.7 & 687 & 18.4 & 1.7 & 0.063 & 8.5; 6.6 & 63; 48\\ 
186 & 3 & 9.0 & 1.1 & 2.7 & 645 & 17.3 & 1.6 & 0.063 & 10.9; 6.6 & 78; 48\\ 
  \enddata
  \tablenotetext{a}{Number of member stars in Fornax sample}
  \tablenotetext{b}{Adopted mass-to-light ratio of the luminous component}
  \tablenotetext{c}{Mass of the dark component.  The first value is the total mass of the dark component; the second value is the dark mass inside r$ \leq 2500$ pc.}
  \tablenotetext{d}{V-band Mass-to-light ratio in solar units.  The first value is the global M/L; the second value is $M/L$ inside r$ \leq 2500$ pc.}
\end{deluxetable*}

\ed